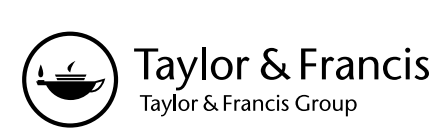

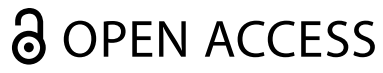

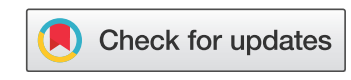

# Hierarchical accompanying and inhibiting patterns on the spatial arrangement of taxi local hotspots

Xiao-Jian Chen[a,b], Quanhua Dong[a,b], Changjiang Xiao[c], Zhou Huang[a], Keli Wang[a], Weiyu Zhang[a] and Yu Liu[a,b]

[a]Institute of Remote Sensing and Geographic Information Systems, School of Earth and Space Sciences, Peking University, Beijing, People's Republic of China; [b]Ordos Research Institute of Energy, Peking University, Inner Mongolia, People's Republic of China; [c]College of Surveying and Geo-Informatics, Tongji University, Shanghai, People's Republic of China

**ABSTRACT**

The spatial arrangement of taxi hotspots indicates their inherent distribution relationships, reflecting their spatial organization structure, and has received attention in urban studies. Previous studies have primarily explored large-scale hotspots through visual analysis or simple indices, which typically spans hundreds or even thousands of meters. However, the spatial arrangement patterns of small-scale hotspots representing specific popular pick-up and drop-off locations have been largely overlooked. In this study, we quantitatively examine the spatial arrangement of local hotspots in Wuhan and Beijing, China, using taxi trajectory data. Local hotspots are small-scale hotspots with the highest density near the center. Their optimal radius is adaptively calculated based on the data, which is 90 m × 90 m and 110 m × 110 m in Wuhan and Beijing, respectively. Popular hotspots are typically surrounded by less popular ones, although regions with many popular hotspots inhibit the presence of less popular ones. These configurations are termed as hierarchical accompanying and inhibiting patterns. Finally, inspired by both patterns, a KNN-based model is developed to describe these relationships and successfully reproduce the spatial distribution of less popular hotspots based on the most popular ones. These insights enhance our understanding of local urban structures and support urban planning.

## 1. Introduction

Taxi pick-up and drop-off hotspots, which are aggregations of stops (Palaniswami et al. 2020), characterize places with intensive activity. Their spatial distribution is closely related to the urban environment (Aslam et al. 2012; Cui et al. 2021; Um and Um 2015), and has attracted attention in traffic management, urban planning, and other fields (Miller et al. 2019; Sila-Nowicka et al. 2016).

**CONTACT** Quanhua Dong dqh@pku.edu.cn Institute of Remote Sensing and Geographic Information Systems, School of Earth and Space Sciences, Peking University, Beijing 100871, People's Republic of China

Although hotspot distribution is influenced by the external urban environment, its internal spatial arrangement pattern is an important topic. Spatial arrangement indicates the spatial relationship between different hotspots and provides information about the structural organization of the space. By checking the distances between hotspots, previous studies have found that hotspots were more densely aggregated in the urban core (Bi et al. 2021b), whereas they were more dispersed on the urban fringe (Bi et al. 2021a). The spatial distribution of Moran's I index for grid-based hotspots further validated this observation, indicating that the urban core consisted of a grid pattern characterized by high-high clustering mixed with a small number of high-low outliers, whereas the pattern in the urban fringe was the opposite, showing low-low clustering with a few low-high outliers (Zhang et al. 2021). Additionally, the popularity of hotspots in the urban core was higher than that in the urban fringe, creating a spatial arrangement in which lower-popularity hotspots surround higher-popularity ones (Nong et al. 2019). Therefore, these findings showed that high-popularity hotspots were concentrated in the urban center, and low-popularity hotspots were scattered in the urban fringe. This indicated that people used the urban core more frequently than the urban periphery, which was consistent with actual experience. By dividing hotspots into more levels based on popularity, it became clear that even within the urban core, a polycentric structure existed, with hotspots of varying intensities distributed in a spatially mixed pattern (Sun and Fan 2021).

These studies have primarily focused on the overall structure of urban spaces by qualitatively exploring large-scale hotspots. First, these hotspots encompassed extensive areas, typically covering entire central business districts (CBDs), train stations, or densely populated residential areas, with radii ranging from hundreds to thousands of meters (Chen et al. 2011; Kumar et al. 2016; Liu et al. 2021). Examination of map scales in these articles further indicated that hotspot coverage extended to several kilometers. Moreover, these studies predominantly investigated patterns through visual analysis or simple indices, often by visually examining the spatial relationships between hotspots or analyzing the distances between them.

Unlike the aforementioned large-scale hotspots, taxi stops exhibit meaningful hotspots on a smaller scale. This is because of people's preference for specific pick-up and drop-off locations (Chang, Tai, and Hsu 2010), such as shopping mall entrances, bus stations, and intersections. The reasonable radius for such hotspots is typically less than 100 m (Chen et al. 2021), considering factors such as the acceptable walking distance when searching for a taxi (Faghih-Imani et al. 2017; Zheng, Liang, and Xu 2012), the size of a popular road in front of a shopping mall (Zhou et al. 2019), and the scale of most buildings (Hulley et al. 2019). These small-scale hotspots reflect how people use urban spaces. Their spatial arrangement patterns can complement local structural information regarding the overall structure of an urban space.

However, the spatial arrangement patterns of small-scale hotspots have not received significant attention. Current studies on small-scale hotspots have primarily focused on detection algorithms and their applications. These investigations extracted small-scale hotspots from larger ones by considering POI distributions (Chen et al. 2020), iteratively applying DBSCAN (Pan et al. 2012), and continuously segmenting hotspots both vertically and horizontally (Chen et al. 2014). Such small-scale hotspots further supported fine-grained land use classification and route recommendations. Recently, Chen et al. (2021) examined the spatial distribution patterns of small-scale hotspots and proposed a detection algorithm based on local density maxima to simulate local pick-up and drop-off preferences. They identified an asymmetric spatial pattern between the pick-up and drop-off hotspots. However, the spatial arrangements of these hotspots have not yet been explored.

This study quantitatively explores the spatial arrangement patterns of small-scale taxi hotspots with varying popularity in Wuhan and Beijing, China. Different levels represent variations in people's activity intensities, thus providing a multilevel characterization of spatial usage. The detection algorithm proposed by Chen et al. (2021) is used to identify small-scale hotspots, referred to by them as 'local hotspots,' a term we will adopt in subsequent research. The hotspots are then classified into multiple popularity levels using the Loubar method (Louail et al. 2014). The main contributions of this study are summarized as follows:

- Compared with previous studies that qualitatively examined the spatial arrangement of large-scale hotspots, we quantitatively study the spatial arrangement of small-scale local hotspots in taxis.
- We demonstrate two interesting hierarchical patterns: Higher popularity hotspots tend to be surrounded by lower popularity hotspots (accompanying pattern); however, in areas with dense higher popularity hotspots, the presence of lower popularity hotspots is inhibited (inhibiting pattern).
- Inspired by these two observed patterns, a KNN-based model is proposed to describe the relations between hotspots with different levels of popularity.

The remainder of this paper is organized as follows. Section 2 introduces the study area and datasets used in this research. In Section 3, the results of the local hotspot identification are described. Section 4 describes the observed accompanying and inhibiting patterns. In Section 5, the KNN-based model is proposed to describe the quantitative relationship between the spatial distributions of local hotspots. Section 6 discusses the relation between local hotspots and points of interest (POIs) as well as applications. Finally, conclusions and future work are presented in Section 7.

## 2. Data sets

Our research utilizes taxi datasets from two major cities in China: Wuhan and Beijing. Pick-up and drop-off locations' latitude and longitude can be extracted per trip. For simplicity, only the results for pick-up stops in Wuhan are presented in the main text (see the other results in the Appendix). The study area is the main urban region, spanning the latitudes 30.467°N to 30.659°N (approximately 21.3 km) and longitudes of 114.181°E to 114.415°E (approximately 22.3 km). The data cover the period from 1 February to 8 August 2015, excluding days of the Chinese New Year (18 February to 25 February 2015) to avoid potential anomalies arising from holiday-related movements. In addition, certain days with limited records, as advised by the taxi company, are excluded because of equipment failure or adverse weather conditions. Consequently, 29,623,951 pairs of pick-up and drop-off stops spanning 163 days are retained for analysis. The analysis is based on a 10-m grid. The spatial density distribution of pick-up stops is shown in Figure 1.

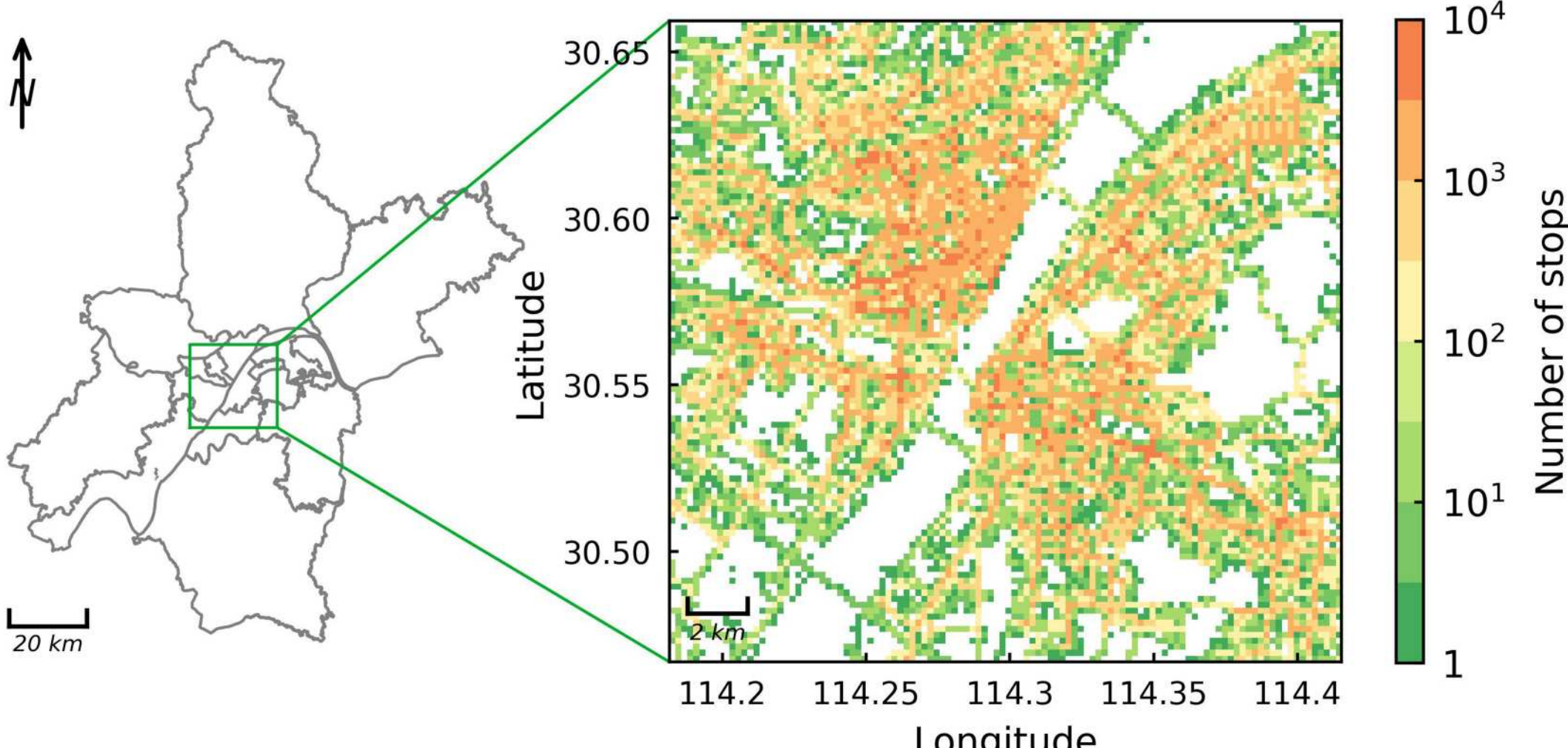

**Figure 1.** Density map of pick-up stops at the 200-m grid scale in the main urban area of Wuhan for visualization.

## 3. Local hotspot identification

### 3.1. Identification method

A local hotspot is a small area with a local maximum density near its center. We focus on this local hotspot for two reasons: (1) It aligns with people's spatial cognition, as individuals tend to get on and off near landmarks, leading to localized density peaks. (2) It offers a comprehensive and detailed characterization of popular pick-up and drop-off locations that exist in both popular and relatively less popular areas.

The local maximum density (LMD) method is grid-based and can be summarized in three steps (Figure 2). Additional details can be found in the original manuscript (Chen et al. 2021).

1) Local maximum determination. It finds the grid with the maximum density among its neighborhood, which is the $(2r+1) \times (2r+1)$ square. The grid radius $r$ can be set manually by the user or adaptively by the 'elbow point' method. The local maximum grid and its neighborhood are collectively referred to as a preliminary local hotspot.

The 'elbow point' method determines the $r$ value where the inclusion of stops in all preliminary local hotspots increases insignificantly as $r$ increases. To achieve this, it first defines the set of local maximum density grids for $r$ as

$$Grid_{LMD}(r) = \{g_{ij} | N_{ij} = \max_{g_{mn} \in E_{ij}(r)} N_{mn}\} \tag{1}$$

where $g_{ij}$ is the grid, $N_{ij}$ is the number of points in $g_{ij}$, $E_{ij}(r)$ is the $(2r+1) \times (2r+1)$ square neighborhood of $g_{ij}$. Then the the grids belonging to squares centered around each $g_{ij} \in Grid_{LMD}(r)$ (i.e. $E_{ij}$) is

$$Grid_E(r) = \{g_{mn} | \exists g_{ij} \in Grid_{LMD}(r),\ s.t.\ g_{mn} \in E_{ij}(r)\} \tag{2}$$

The number of covered stops is

$$N(r) = \sum_{g_{ij} \in Grid_E(r)} N_{ij} \tag{3}$$

Finally, to describe the 'elbow point' mathematically, $r$ represents the minimum value at which the second deviation of $N(r)$ reaches a local minimum (Yuan and Raubal 2014).

2) Neighborhood reshaping. After the first step, some neighborhoods may overlap with each other. The overlapping grids require a unique class to a local maximum grid. The gravity rule is then used to calculate the attraction of a central local maximum grid $g_{i_1,j_1}$ to an overlapping grid $g_{i_0,j_0}$ as

$$G(g_{i_1,j_1} \rightarrow g_{i_0,j_0}) = \frac{N_{g_{i_1,j_1}}}{d(g_{i_1,j_1}, g_{i_0,j_0})} \tag{4}$$

where $N_{g_{i_1,j_1}}$ is the number of stops in grid $g_{i_1,j_1}$ and $d(g_{i_1,j_1}, g_{i_0,j_0})$ is the Euclidian distance between the centers of the two grids. Similarly, we can calculate $G(g_{i_2,j_2} \rightarrow g_{i_0,j_0})$ and class $g_{i_0,j_0}$ to the one with the largest attraction.

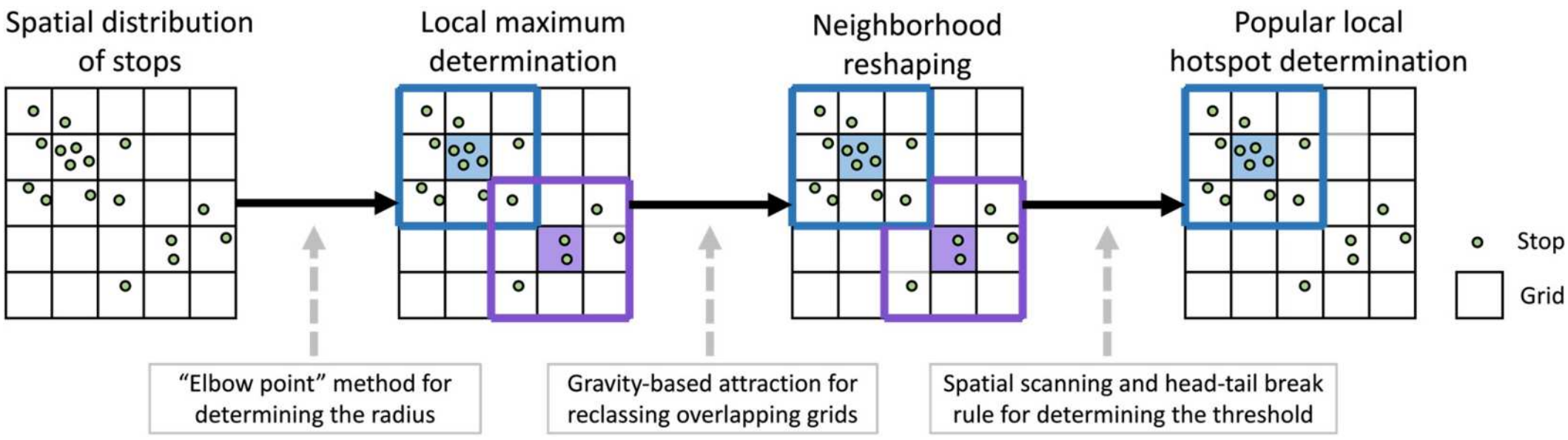

**Figure 2.** Flowchart of local hotspots identification.

3) Popular local hotspot determination. The final step is to determine whether a non-overlapping local hotpot is sufficiently popular, using a threshold. To adaptively set this threshold, a spatial scanning method is adopted, that is, randomly selecting neighborhoods with the same shape as the local hotspot, with a total of 10,000, $N_{r_1}, \ldots, N_{r_{10000}}$. The head–tail break rule (Jiang 2013) is repeatedly applied until the remaining count is less than 4000 (i.e. 40% of 10,000). The breaking value at this stage is considered the threshold for the local hotspot.

In the subsequent analysis, we refer to the local maximum grid of each local hotspot as the 'center' and the distance between two local hotspots as the Euclidean distance between their centers.

### 3.2. Spatial distribution of local hotspots and level classification

Figure 3 illustrates how the 'elbow point' method is used to determine the grid radius $r$ for identifying taxi pick-up stops in Wuhan. As shown, the number of covering stops, $N(r)$, increases as the grid radius increases. The first local minimum of the second derivative of $N(r)$ occurres at $r = 4$, indicating the point at which the growth of $N(r)$ begin to decelerate significantly. Given that the grid size is 10 m, the final square neighborhood is defined as 90 m × 90 m (i.e. 9 × 9 grids). It is noteworthy that different datasets have their own optimal grid radii. In the case of Beijing, $r = 5$ (Appendix B), which corresponds to a square neighborhood of 110 m × 110 m (i.e. 11 × 11 grids).

Figure 4 shows the identified local hotspots in a CBD and an ordinary street. The CBD, known as 'Jiefang Road,' serves as a prime location for leisure and entertainment (Figure 4a). Using Step 1, square neighborhoods of 90 m × 90 m are identified, with the local maximum density in the center (Figure 4b). After undergoing steps 2 and 3, the square neighborhoods are transformed into non-overlapping ones, and only those containing sufficient stops are retained (Figure 4c). These hotspots are found near intersections and typical popular buildings. For instance, on the southwest side, there are K11 Shopping Center and Xianghe Apartment. K11 Shopping Center offers luxury goods and high-end culinary establishments. It also hosts art exhibitions throughout the year. Xianghe Apartment, on the other hand, is an upscale residential area. The northwest side features

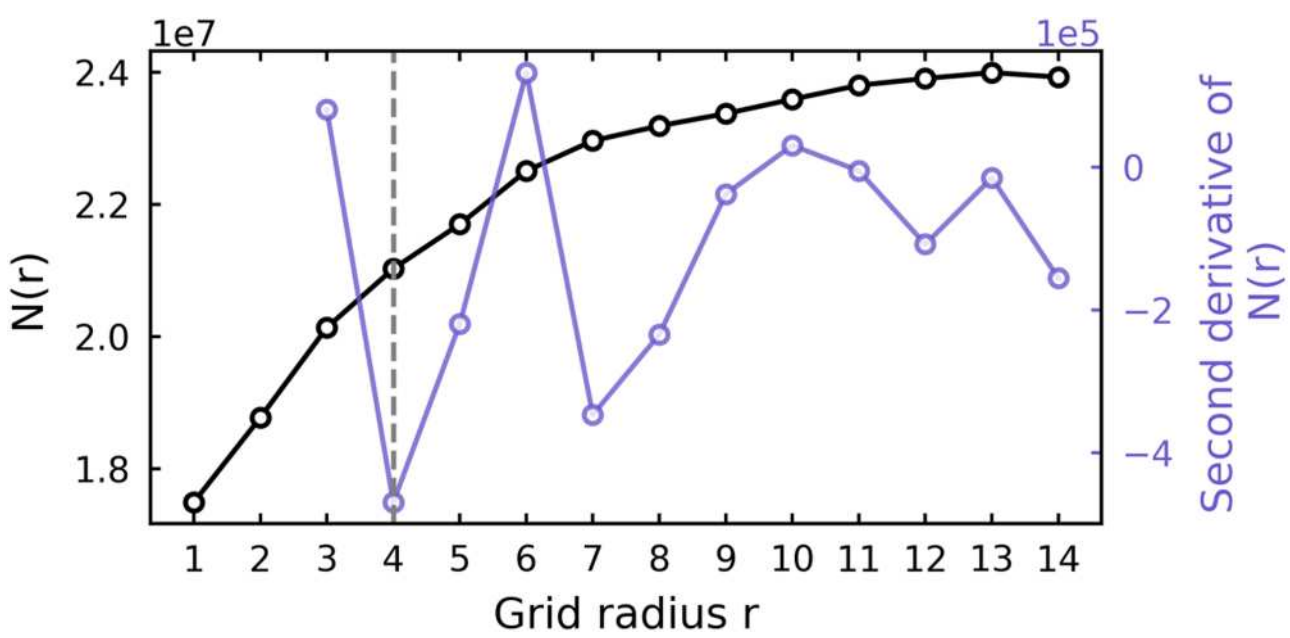

**Figure 3.** Values of $N(r)$ and $N''(r)$ as the grid radius r increases for pick-up stops in Wuhan. According to the 'elbow point' method, $r$ is equal to 4.

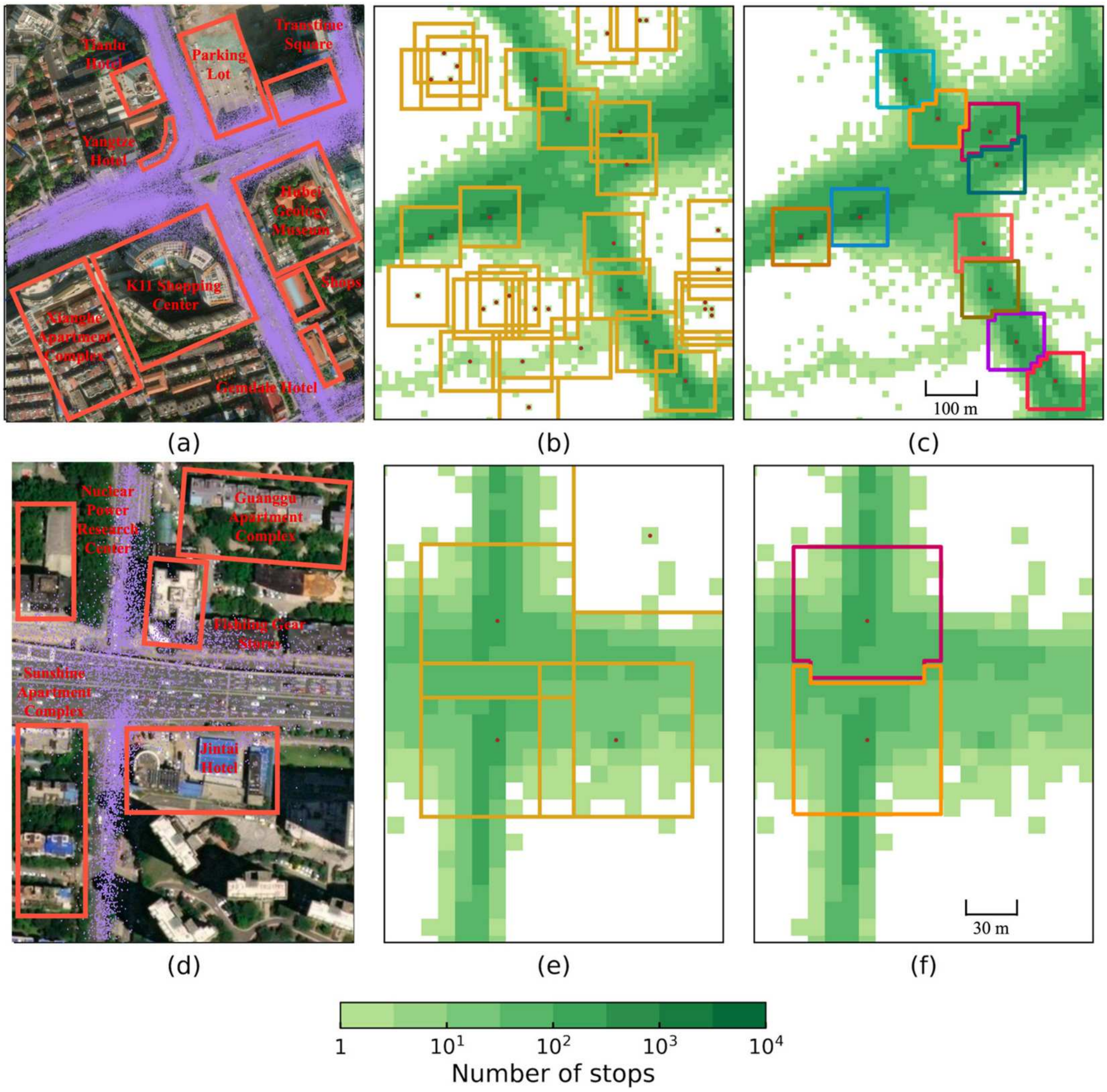

**Figure 4.** Examples of the local hotspots of taxi pick-up stops in a CBD (a–c) and an ordinary street (d–f) in Wuhan. In (a) and (d), purple dots indicate observed taxi stops. In (b) and (e), the orange square indicates a preliminary local hotspot with the local maximum density in the center (red dot). In (c) and (f), each polygon represents the final detected local hotspot.

Tianlu and Yangtze Hotels, known for being boutique hotels in the bustling city center, favored by businesspeople. Moreover, the southeast houses the Hubei Geology Museum, attracting a large number of tourists daily. The Transtime Square, located on the northeast side, is a prestigious residential area within Wuhan city.

On an ordinary street (Figure 4d), the main visitors are typically employees of the Nuclear Power Research Center, a small group of individuals purchasing fishing gear, and customers staying at the Jintai Hotel. Local hotspots in this area are primarily found at the intersections and entrances of the three buildings (Figure 4e,f). This suggests that both the popular CBD and the relatively less popular ordinary street exhibit local hotspots near significant POIs, reflecting people's preferences for specific pick-up locations.

A total of 2158 pick-up hotspots with 16,683,351 stops are identified. The Loubar (Bassolas et al. 2019; Louail et al. 2014) method is used to classify the hotspots into multiple levels based on their popularity. The Loubar method is based on the derivation of the Lorenz curve. The Lorenz curve represents the fraction of hotspots versus the fraction of total stops in these hotspots, where the

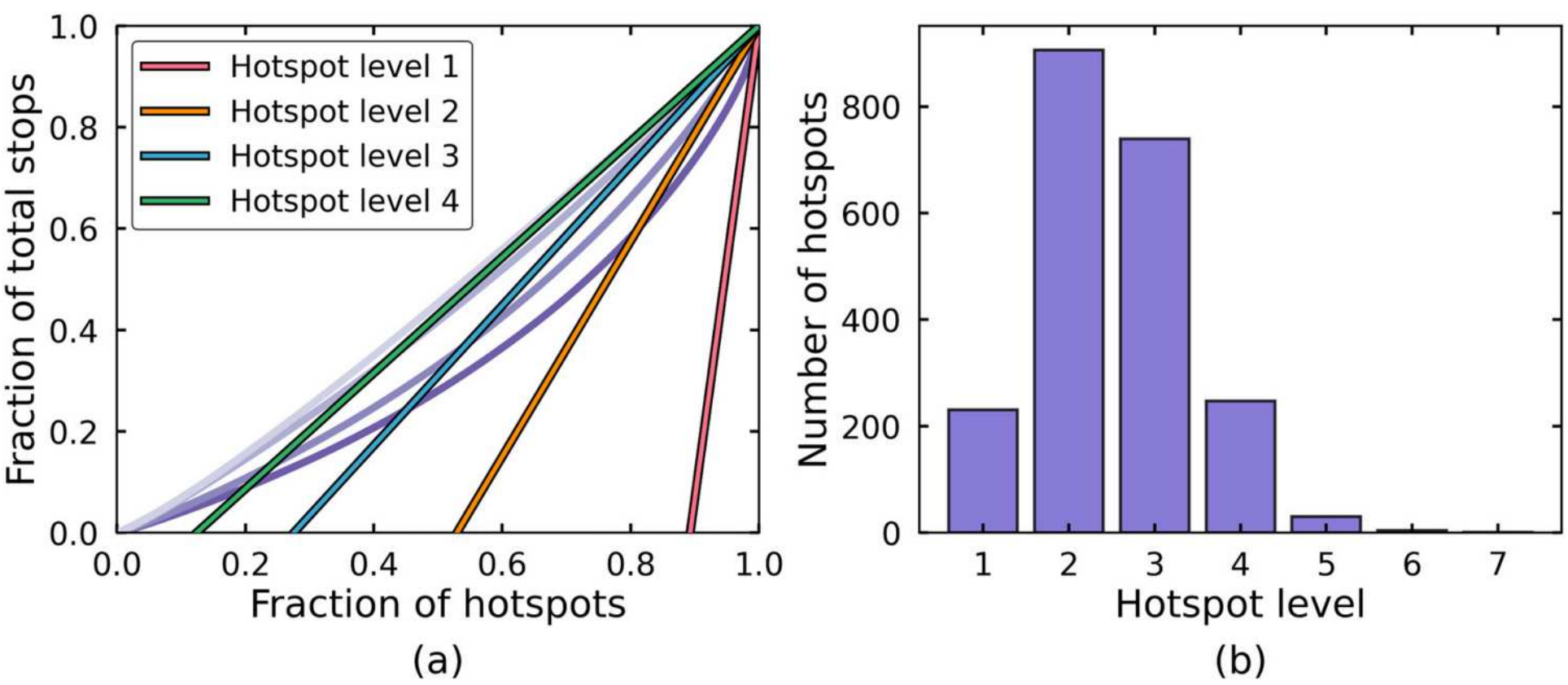

**Figure 5.** (a) Illustrative diagram of calculation for four levels of hotspots using the Loubar method. (b) Number of hotspots in each level.

hotspots are in ascending order. The threshold value is obtained by determining the derivative of the Lorenz curve at (1, 1) and extending it until it intersects the x-axis. Once hotspots are identified at a certain level, they are excluded from the data. The threshold is then recalculated using the new

**Table 1.** The statistical results of different levels of pick-up local hotspots classified by Loubar method.

| Level | Number of hotspots | Range value of fractions of total stops | Range value of stops (median) |
|---|---|---|---|
| 1 | 231 | [0,27.47%] | 72,175–13,357 (16,676) |
| 2 | 906 | [27.47%,74.08%] | 13,346–5847 (8082) |
| 3 | 739 | [74.08%,94.57%] | 5846–3671 (4526) |
| 4 | 247 | [94.57%,99.52%] | 3669–2692 (3413) |
| 5 | 30 | [99.52%,99.95%] | 2680–1981 (2413) |
| 6 | 4 | [99.95%,99.99%] | 1969–1863 (1901) |
| 7 | 1 | [99.99%,100%] | 1490 |

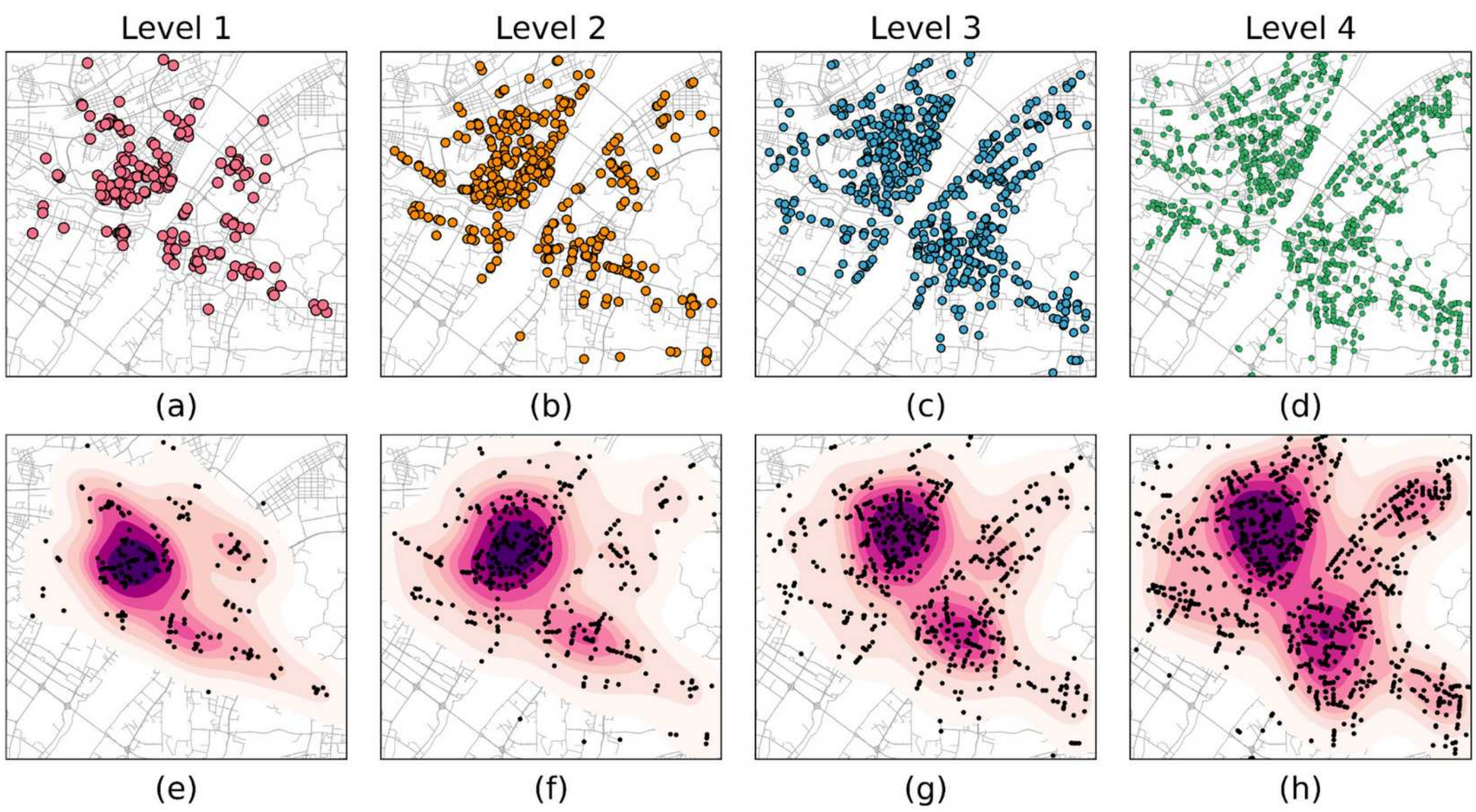

**Figure 6.** Spatial distributions of pick-up local hotspots (a–d) in Wuhan with the kernel density estimation for visualization purposes (e–h).

Lorenz curve for the remaining data. This procedure is conducted iteratively until all the hotspots are classified. Figure 5a demonstrates the process until level 4, where the Lorenz curves are shown in varying levels of transparency in purple shades as the levels descend. The Loubar method identifies seven hotspot levels (Figure 5b), with most stops contained in the first four levels. The first four levels had 2123 of the 2158 hotspots, representing 99.52% of the stops (Table 1). Therefore, we focus on the results of the first four levels.

Their spatial distributions display two interesting patterns (Figure 6). First, higher-level hotspots tend to be surrounded by lower-level hotspots. This also results in very similar spatial distributions of different-level hotspots. Second, although the spatial distributions are similar, the lower-level hotspots are not always clustered near the higher-level hotspots, but expand outward to a certain extent. In the following section, we demonstrate that the presence of lower-level hotspots is inhibited in regions where higher-level hotspots are densely distributed.

## 4. Accompanying and inhibiting patterns for empirical analysis

This section describes the accompanying and inhibiting hierarchical patterns. The methods for the characterization of these two patterns are introduced in Section 4.1, followed by the empirical results in Section 4.2.

### 4.1. Approaches for characterization

#### 4.1.1. Characterization of accompanying pattern

We call the phenomenon that higher-level hotspots are likely to have nearby lower-level hotspots the 'accompanying pattern'. Measuring spatial proximity is the key to proving this. To aid in this, we introduce the k-nearest neighbor (KNN) distance and coverage ratio. The KNN distance measures the proximity of lower-level hotspots adjacent to high-level hotspots, whereas the coverage ratio describes the proportion of low-level hotspots within a given radius of high-lever hotspots.

The KNN distance (Elhorst, Lacombe, and Piras 2012) refers to the distance between hotspot $x$ and the $k$ nearest neighbor hotspot of set $B$. Then, the average KNN distance of the two sets measures the spatial nearest neighbor relationship between them, defined as:

$$\overline{KNN}(A, B, k) = \frac{1}{|A|}\sum_{x \in A} rank(k|d(x, y), y \in B) \tag{5}$$

where $A$ and $B$ are the higher-level and lower-level hotspots sets, $rank(k|d(x, y), y \in B)$ means the $k$-th nearest for the distances from $x$ to the hotspots in $B$. The smaller $\overline{KNN}$ is the closer it is from $A$ to $B$.

Another index is the coverage ratio of higher-level hotspots ($A$) to the next lower-level hotspot ($B$). It is calculated by

$$CR(A, B, r) = \frac{|\{x \in B | d(A, x) < r\}|}{|B|} \tag{6}$$

where $d(A, x)$ is the distance from $x$ to the nearest hotspot in set $A$. In this way, $CR(A, B, r)$ describes the proportion of the next lower-level hotspot $B$ within a radius $r$ of hotspot $A$. The faster $CR(A, B, r)$ increases with $r$, the closer $A$ is to $B$.

Furthermore, to prove the existence of the accompanying pattern, two types of random procedures are conducted as null models to determine whether the observed distances are close. The first one is partial random (named 'random 1'). For a given higher-level hotspot $A_{obs}$ (e.g. level 1) and its next lower-level hotspot $B_{obs}$ (e.g. level 2), it randomly chooses grids on the road as the centers of the lower-level hotspots, recorded as $B_{ran_1}$. Subsequently, $\overline{KNN}(A_{obs}, B_{ran_1}, k)$ and $CR(A_{obs}, B_{ran_1}, r)$ arere calculated to demonstrate the spatial proximity of the observed higher-level and random lower-level hotspots. The second one is completed random (named 'random 2'). It randomly chooses grids on the road for all hotspots and then calculates $\overline{KNN}(A_{ran_2}, B_{ran_2}, k)$ and $CR(A_{ran_2}, B_{ran_2}, r)$. Therefore, by comparing these values with $\overline{KNN}(A_{obs}, B_{obs}, k)$ and $CR(A_{obs}, B_{obs}, r)$, it enables a quantitative description of the difference between observation and random to verify the accompanying pattern.

#### 4.1.2. Characterization of inhibiting pattern

We call the phenomenon that regions where higher-level hotspots are denser and the existence of lower-level hotspots is inhibited to a certain extent the 'inhibiting pattern'. The term 'inhibiting pattern' is borrowed from the field of spatial point process. In this field, the existence of a point reduces the probability of the existence of another point nearby, which is called inhibition (Baddeley, Bárány, and Schneider 2007). Our study emphasizes that only regions with dense (not sparse) higher-level hotspots will reduce the probability of the existence of high-level hotspots, which is different from the description in spatial point process. To verify this conjecture, the spatial density of hotspots at each level is measured. Counting the number of points within a distance $d_{count}$ is a useful metric that is applied in many widely used measures such as Ripley's K-function (Illian et al. 2008).

Formally, we have

$$Count(x, B, d_{count}) = |\{y \in B|\ d(x, y) < d_{count}\}| \tag{7}$$

$$Den_{nor}(x, B, d_{count}|A) = \frac{Count(x, B, d_{count})}{\max\limits_{x \in A} Count(x, B, d_{count})} \tag{8}$$

where $d_{count}$ is the given distance, $A$ and $B$ are two hotspots sets, $x \in A$ is a hotspot. Therefore, $Count(x, B, d_{count})$ represents the number of $B$'s hotspots contained within the $d_{count}$ radius of $x$. $Den_{nor}(x, B, d_{count}|A) \in [0, 1]$ represents the relative number of hotspots (called 'normalized density') comparing to other hotspots in $A$. Particularly, $Den_{nor}(x, B, d_{count}|A) = 1$ denotes $x$ is in the densest location of the $B$.

Let $A$ and $B$ be the higher-level and lower-level hotspot sets, respectively; $Den_{nor}(x, A, d_{count}|A)$ and $Den_{nor}(x, B, d_{count}|A)$ denote the normalized density of $x$ at the same level and the next lower level, respectively. Different $d_{count}$ define the normalized density differently. The inhibiting pattern holds when $Den_{nor}(x, A, d_{count}|A)$ is large and $Den_{nor}(x, A, d_{count}|A) > Den_{nor}(x, B, d_{count}|A)$ for different $d_{count}$.

Figure 7 shows a calculation example for $Den_{nor}(x, A, d_{count}|A)$ and $Den_{nor}(x, B, d_{count}|A)$, where $d_{count}$ is a preset parameter. In this space, the four distant regions each contain varying numbers of higher- and lower-level hotspots. From the figure, it is easy to determine that $\max_{x \in A} Count(x, A, d_{count}) = 3$ and $\max_{x \in A} Count(x, B, d_{count}) = 4$. Then, according to formula (8), $(Den_{nor}(x, A, d_{count}|A), Den_{nor}(x, B, d_{count}|A))$ of the four higher-level hotspots in the center of each dashed circle can be calculated. The first higher-level hotspot contains three higher-level (including itself) and four lower-level hotspots within the radius $d_{count}$, leading to the normalized densities of both being equal to one. In other words, it is located in an area with the highest density of both higher- and lower-level hotspots. The second hotspot has a slightly lower relative density of surrounding higher-level hotspots. The third type has fewer nearby higher- and lower-level hotspots. Conversely, the fourth is located in the most densely populated area for higher-level hotspots but with only one lower-level hotspot nearby, demonstrating an inhibiting pattern.

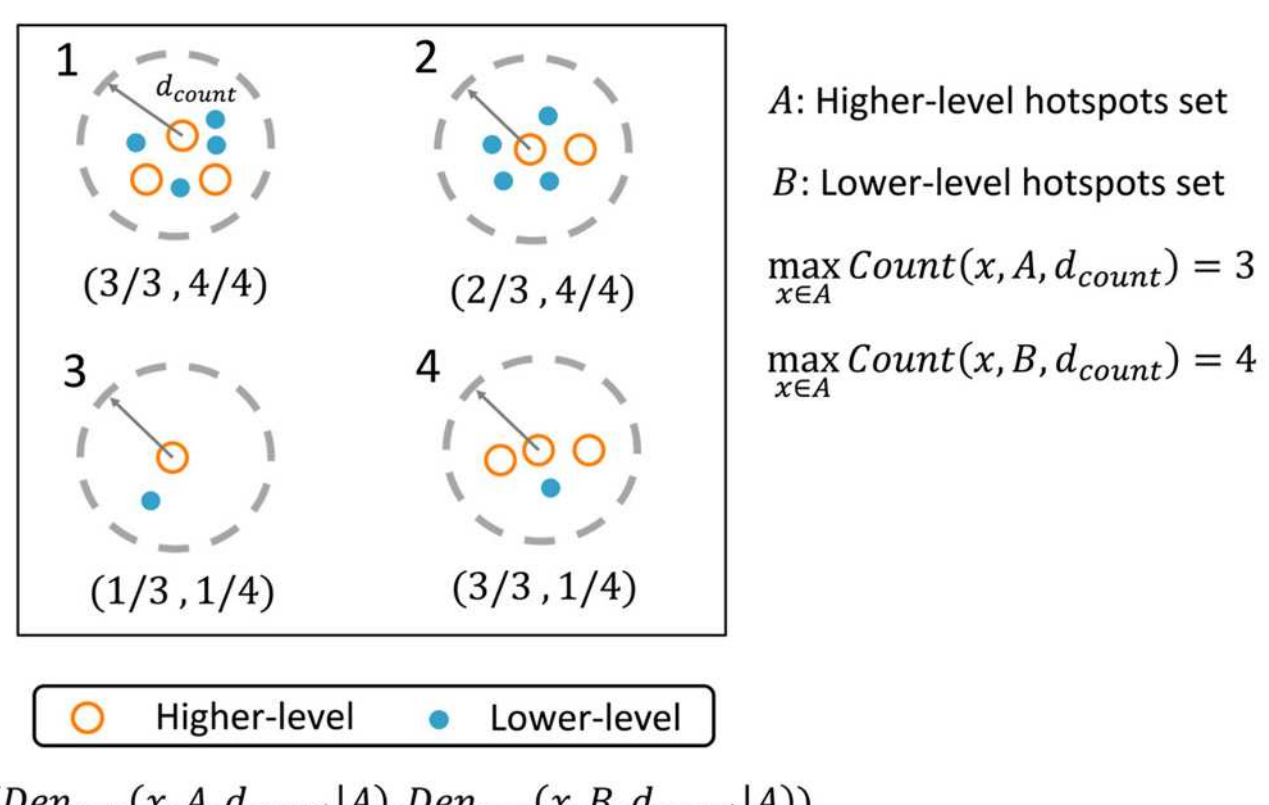

**Figure 7.** Calculation examples for $Den_{nor}(x, A, d_{count}|A)$ and $Den_{nor}(x, B, d_{count}|A)$.

## 4.2. Empirical result

### 4.2.1. Accompanying pattern

Figure 8a–c show the $\overline{KNN}(A, B, k)$ curves between hotspots and their subsequent lower-level hotspots. Compared with the 100 random results, the observed values are significantly smaller. Specifically, the average nearest distances of hotspots for these three levels are 169.08, 266.36, and 478.16 m, indicating that a hotspot can find a low-level hotspot, on average, at this distance. For the coverage ratio, the observed $CR(A, B, r)$ increases faster than the random curves (Figure 8d–f). Therefore, both results indicate that the observed lower-level hotspots are closer to the higher-level hotspots than would be the case by random, which proves the accompanying pattern.

### 4.2.2. Inhibiting pattern

Figure 9a–c show the pair results of $(Den_{nor}(x, A, d_{count}|A), Den_{nor}(x, B, d_{count}|A))$ for $d_{count} = 1000$ m. While the mean curves show an overall increase in volatility, they mostly fall below the curve $y = x$ at a larger $Den_{nor}(x, A, d_{count}|A)$. This implies that in the denser areas of higher-level hotspots

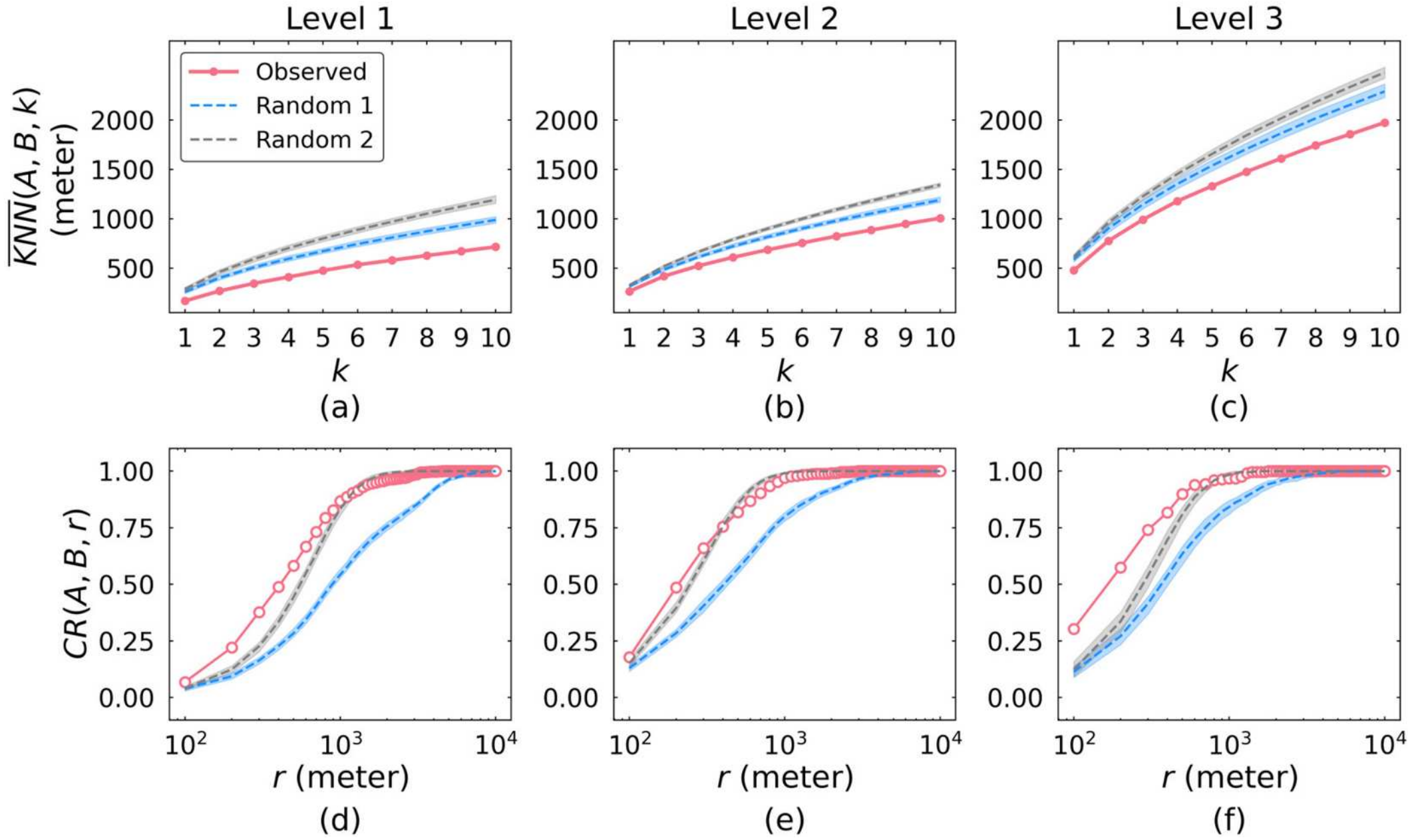

**Figure 8.** (a–c) $\overline{KNN}(A, B, k)$ and (d–f) $CR(A, B, r)$ at each level, where $A$ and $B$ are the higher-level and its next lower-level hotspot sets, respectively. In each subfigure, the dashed line represents the median value for random procedures, and the shaded area corresponds to the 10–90% quantile range.

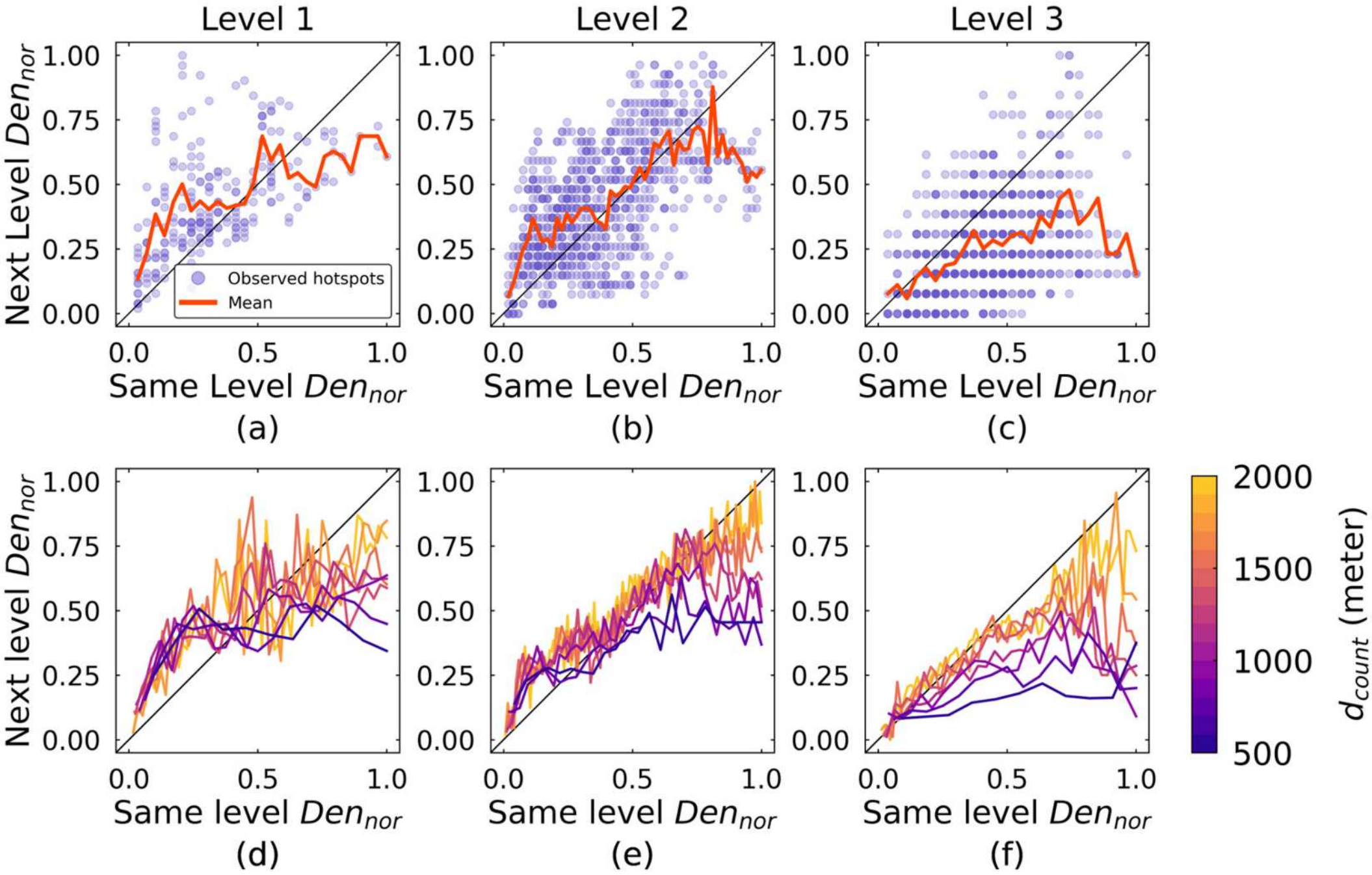

**Figure 9.** (a–c) Scatterplots of ($Den_{nor}(x, A, d_{count}|A)$, $Den_{nor}(x, B, d_{count}|A)$) with $d_{count} = 1000$ m, where $x \in A$, $A$ and $B$ are the sets of hotspots in a level and the next lower-level. In each subplot, the red line connects the mean of $Den_{nor}(x, B, d_{count}|A)$s for each $Den_{nor}(x, A, d_{count}|A)$. (d–f) Mean curves with different $d_{count}$.

(i.e. $Den_{nor}(x, A, d_{count}|A)$ is larger), the relative density of lower-level hotspots is relatively low (i.e. $Den_{nor}(x, B, d_{count}|A) < Den_{nor}(x, A, d_{count}|A)$). In general, the mean curves with different $d_{count}$ exhibit a similar trend, which further verifies the inhibiting pattern (Figure 9d–f).

To provide an intuitive description, Figure 10 shows the spatial distribution of different $Den_{nor}$ values with $d_{count} = 1000$ m. Each column shows the spatial distribution of hotspots at each level. The color of the hotspots in the first row represents the normalized density at the same level (i.e. $Den_{nor}(x, A, d_{count}|A)$), whereas the color in the second row represents the density at the next lower level (i.e. $Den_{nor}(x, B, d_{count}|A)$). Brighter colors indicate higher densities. The color in the third row represents $Den_{nor}(x, B, d_{count}|A) - Den_{nor}(x, A, d_{count}|A)$. Lower values (green) indicate that the next level is less dense than the current level. The three dashed circles (in Figure 10) show examples of regions that exhibit inhibiting pattern. Both regions had denser same-level hotspots but relatively less dense lower-level hotspots. This leads to a relatively small value of $Den_{nor}(x, B, d_{count}|A) - Den_{nor}(x, A, d_{count}|A)$ (i.e. darker green in the last row).

Figure 11 shows maps of the three regions. These areas include the CBD, large hospitals, and popular tourist attractions, drawing many daily visitors. The center of Region 1 is approximately at Tongji Hospital. Tongji Hospital is the largest hospital in Wuhan, with its outpatient and inpatient departments located together on a large campus. It serves a substantial number of patients daily. Adjacent to Tongji Hospital is the K11 shopping center, which attracts many shoppers each day. The center of Region 2 is situated around Yongtai shopping center, a large retail complex near Wuhan River Beach, one of the city's most famous attractions. These sites draw numerous local residents and tourists daily. The center of Region 3 is located near Zhongshan Park, one of Wuhan's largest parks. Near Zhongshan Park is another prominent hospital, Xiehe Hospital. Interestingly, all three regions feature four levels of hotspots, which are spatially intermingled. Hotspots of levels 1 and 2 are primarily located along main roads, while levels 3 and 4 are found on smaller streets.

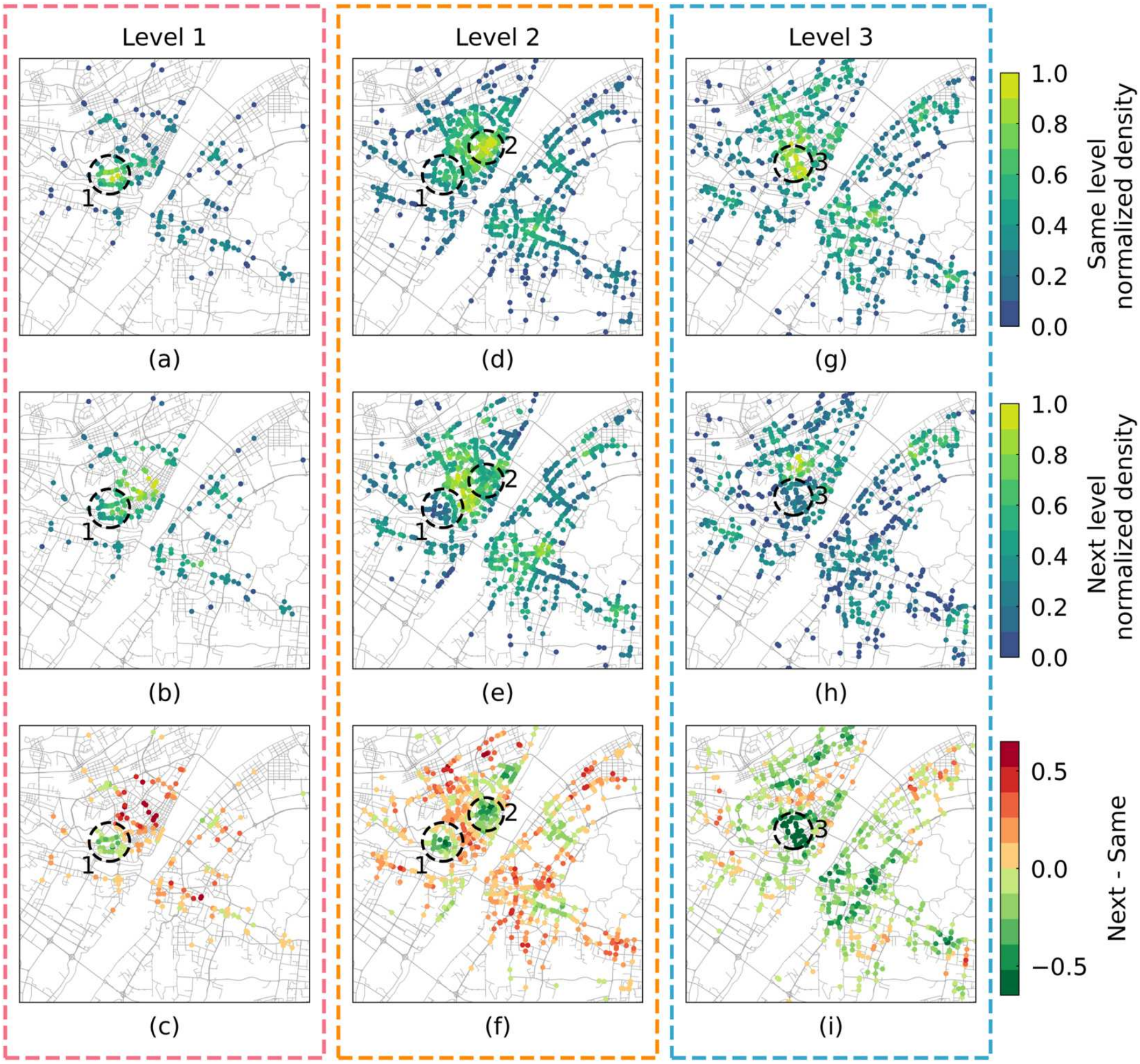

**Figure 10.** Spatial distribution of pick-up hotspots with different levels and normalized density labeled by colors. Noticing that 'next lower-level' is simplified as 'next level' in this figure.

To summarize, to show the accompanying and inhibiting patterns intuitively, Figure 12 illustrates three toy examples of local hotspot distribution, where the spatial distribution of higher-level hotspots is provided in advance. The numbers of higher- and lower- level hotspots are seven and fourteen, respectively.

- As shown in Figure 12a, in the random case, the lower-level hotspots are relatively evenly dispersed in the space and are not affected by the higher-level hotspots.
- As shown in Figure 12b, in the case of only the accompanying pattern, two extremely close lower-level hotspots may exist near each higher-level hotspot. Taking the higher-level hotspot in the figure as the center, the number of hotspots within a radius $d$ (i.e. within the dashed circle) is counted. Both higher- and lower-level hotspots account for 4/7 of their respective hotspot numbers. This shows that high- and low-level hotspots are equally dense within the dashed circles.
- As shown in Figure 12c, if the inhibiting pattern is added on the basis of accompanying, the lower-level hotspots do not consistently gather in the dashed circle. Specifically, the proportion of low-level hotspots reduced to 4/14, which is smaller than the relative proportion of higher-level hotspots. Lower-level hotspots still tend to gather toward high-level hotspots; however, in areas with high concentrations, the relative density of low-level hotspots decreases.

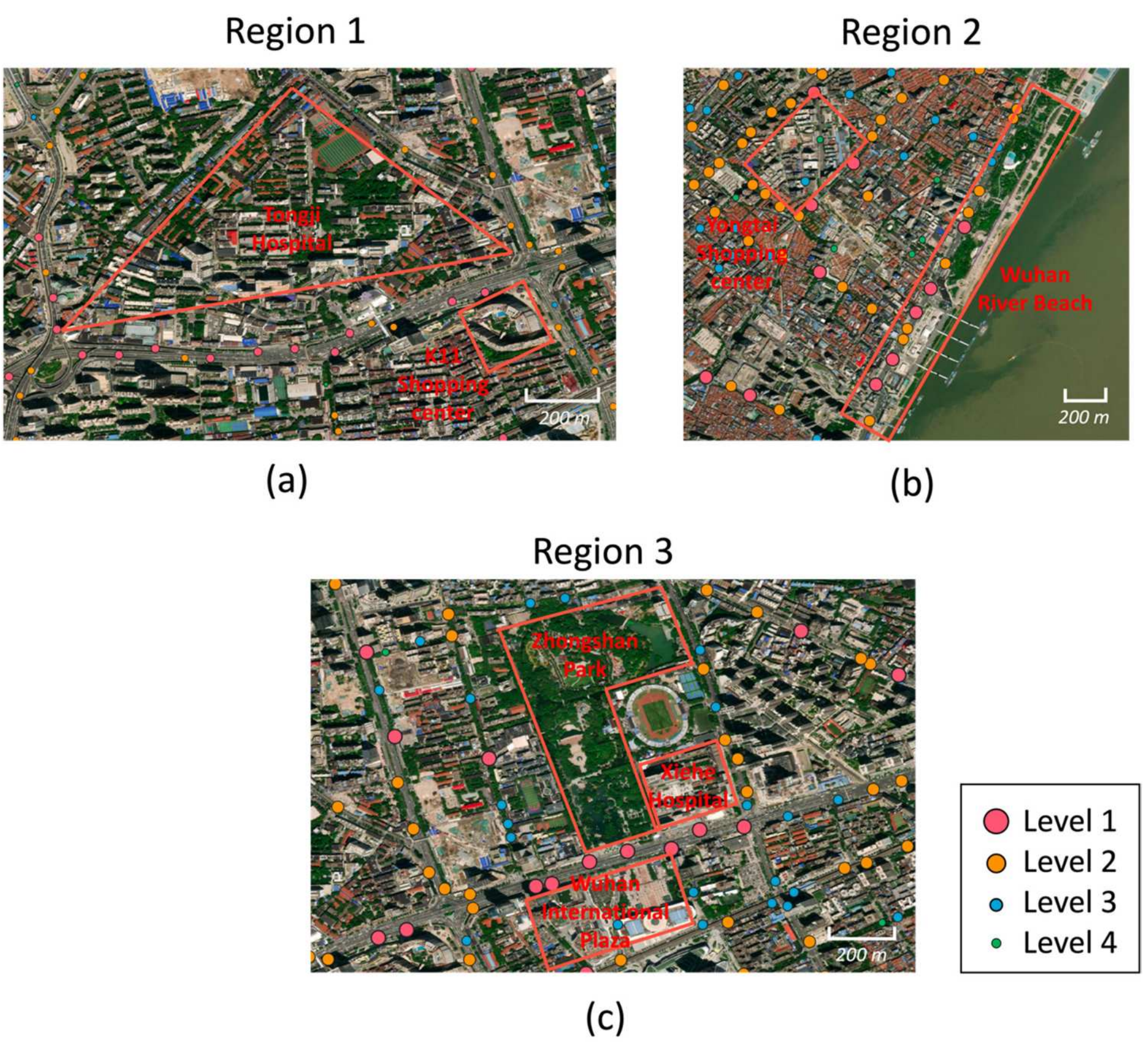

**Figure 11.** Maps of three typical regions circled in Figure 10.

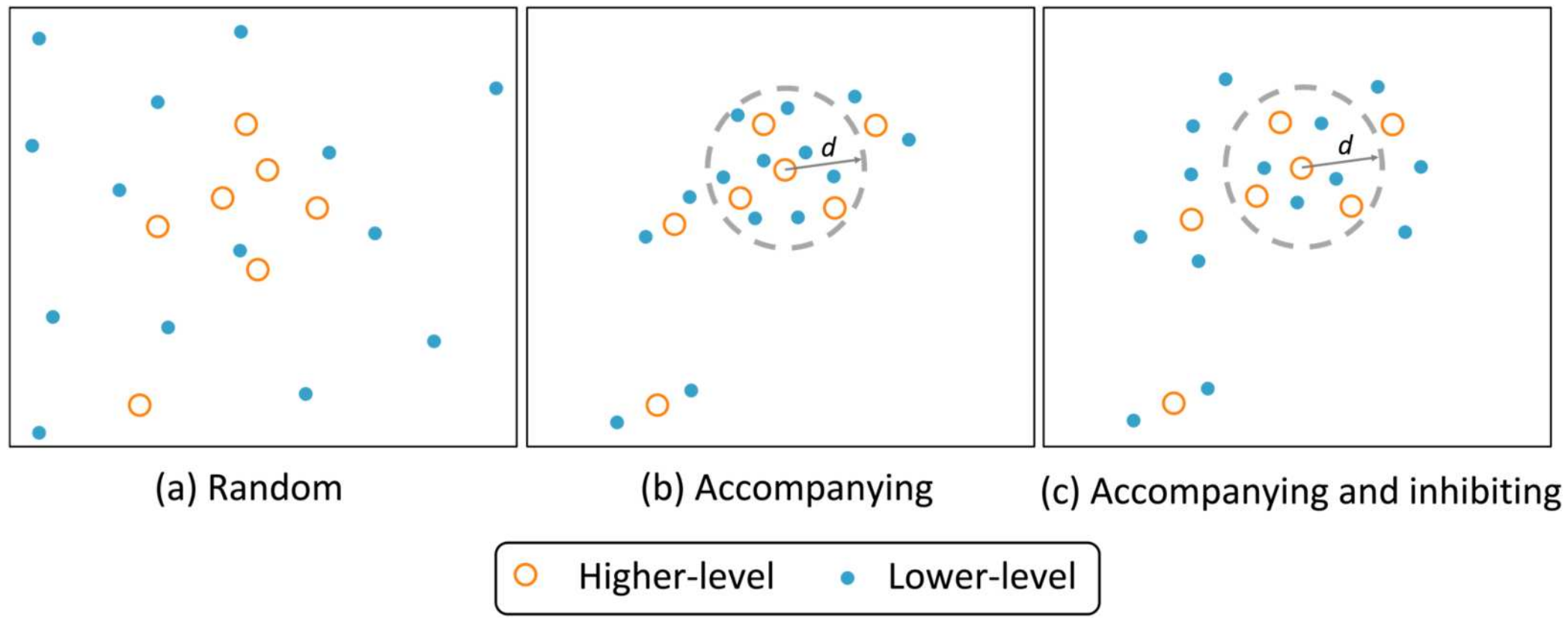

**Figure 12.** Toy example of the spatial arrangement of local hotspots. (a) Randomly distributed lower-level local hotspots. (b) A kind of distribution of lower-level hotspots with the accompanying pattern. (c) Distribution of lower-level hotspots with the accompanying and inhibiting patterns. The dashed circle indicates the region with dense higher-level hotspots.

## 5. KNN-based model for specific relation

We have shown that hotspots exhibit hierarchical accompanying and inhibiting patterns. Another question is whether a specific quantitative relationship exists. In this section, inspired by the two aforementioned patterns, a KNN-based model is proposed to describe this relationship. To verify the correctness of the model, we use the KNN-based model to generate the spatial distributions of other hotspots based on Level 1 hotspots.

### 5.1. Model description

Given a higher-level hotpot set $A$, the probability that a next lower-level hotpot exists at each location $x$ (i.e. the center of the 10 m × 10 m grid) is

$$p(x|A) \sim \sum_{1 \le i \le K} d_{(i)}(x, A)^{-\alpha} \mathbb{1}_{\{d_{(i)}(x,A) \le d_{cut}\}} \quad (9)$$

where $d_{(i)}(x, A)$ is the $i$-th closest distance from the grid $x$ to $A$, $\mathbb{1}$ is the indicator function, $K$, $\alpha$, and $d_{cut}$ are predefined positive parameters. In appendix, different values of the three parameters are used to show the better performances of KNN-based mechanism than other two mechanisms below (i.e. global contribution-based and completely random mechanisms).

Equation (7) suggests that the probability of a location existing in a lower-level hotspot is affected by the cumulative distance-decay effect of the $K$ nearest higher-level hotspots. This simple KNN-based mechanism has also been widely used in geo-related problems such as groundwater level infill (He et al. 2020), POI recommendations (Li et al. 2022), and traffic frequency prediction (May et al. 2008).

This KNN-based mechanism is inspired by accompanying and inhibiting patterns. First, the accompanying pattern indicates that the closer the distance to higher-level hotspots, the larger the probability of the existence of lower-level hotspots. The factor $d_{(i)}(x, A)^{-\alpha}$ ensures that the mechanism aligns with this pattern. Second, the inhibiting pattern indicates that an area with a dense distribution of high-level hotspots will not produce similarly dense low-level hotspots. Low-level hotspots tend to spread outward. The influence of $K$ nearest hotspots guarantees this pattern to a certain extent. This indicates that higher-level hotspots outside the $K$ nearest hotspots of a location do not contribute to the probability of that location having low-level hotspots, resulting in low-level hotspots being less dense than high-level ones.

To demonstrate the consistency between the KNN-based model and two patterns intuitively, Figure 13 presents an example of spatial probability by six higher-level hotspots with $K = 2$ and $\alpha = 1$ in a Euclidean space. First, the high probability around hotspot centers causes lower-level hotspots to be distributed around them, in line with the accompanying pattern. Second, the probability of the inner area being surrounded by the six local hotspots is not significantly higher than that of the outer area. For example, grid C is located further outside than grid D, but both have the same probability, and the area located in the dashed circle has a low probability, even though it is surrounded by four higher-level hotspots. Therefore, this mechanism guarantees that lower-level hotspots are not always concentrated in the inner area and appear to spread outward. This is consistent with the inhibiting pattern.

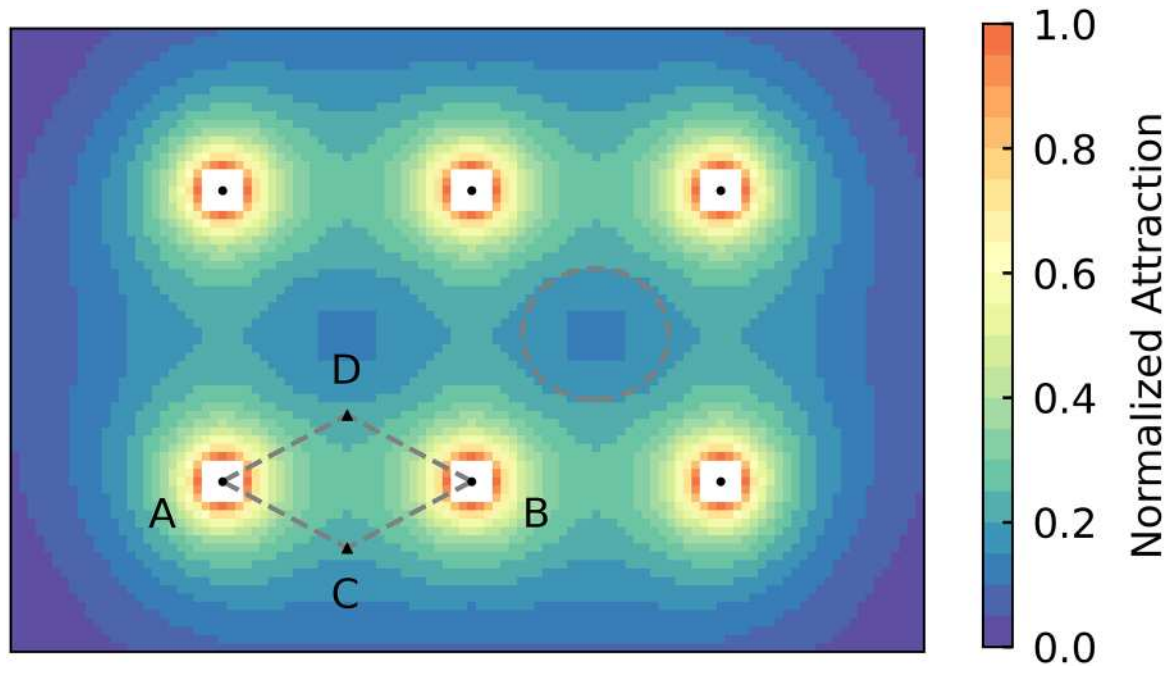

**Figure 13.** Example of the spatial distribution of min-max normalized probability generated by six high-level hotspots, where $K = 2$, $\alpha = 1$. The grids within 4 grid distances (40 m) around the central grids (black circle dots) of the hotspot do not have normalized attractiveness and are displayed in white color.

Notably, the KNN model only simulates the location of the local hotspot center without considering the accurate boundaries between local hotspots. Specifically, in the simulation process of selecting the low-level hotspot center, we exclude the existing hotspots and their square neighborhoods with a grid radius of 4, and then selected a grid as the new center in each iteration (see Appendix A). For example, an area with a grid radius of four around each black circular dot in Figure 13 cannot be selected (i.e. the white area in the figure). However, we did not further simulate accurate boundaries between hotspots. This is a limitation of the model.

To justify the KNN-based mechanism, we compare two other mechanisms: global contribution-based and completely random. The differences are in the attraction functions:

(1) Global contribution-based

$$p(x|A) \sim \sum_{y \in A} d(x, y)^{-\alpha} \mathbb{1}_{\{d(x,y) \leq d_{cut}\}} \tag{10}$$

(2) Completely random

$$p(x|A) \sim \mathbb{1}_{\{d_1(x,A) \leq d_{cut}\}} \tag{11}$$

Thus, within $d_{cut}$, the global contribution-based mechanism indicates lower-level hotspots are influenced by all surrounding higher-level hotspots, and the completely random mechanism means higher-level hotspots are not attractive to lower-level ones.

### 5.2. Simulation problem setting

A simulation is established to verify the accuracy of the models. Based on the Level 1 hotspots, we use three models to reproduce the spatial distribution of the Level 2 hotspots. Based on the simulated Level 2 hotspots, Level 3 hotspots are then reproduced. This procedure continues until the hotspots of all levels are reproduced. Note that the simulation is the grid location where the center of the local hotspot is situated.

In our simulation, the relations are restricted to a range $d_{cut}$, i.e. a higher-level hotspot only influences the $d_{cut}$ range. Because a local hotspot represents a specific popular place, it is unreasonable to assume a relationship between higher- and lower-level hotspots, regardless of how far away they are. This setting also leads to slight differences in the simulation problem. The next lower-level hotspots outside the $d_{cut}$ of the higher-level hotspots are retained at the beginning because they could not be reproduced. For a detailed discussion on $d_{cut}$, please refer to the Appendix C.

To measure the spatial similarity between the observed and simulated local hotspots, recorded as $O$ and $R$, the following pair of results is used to calculate the RMSE: Specifically,

$$Com(d_{RMSE}) = \{(Count(x, O, d_{RMSE}), Count(x, R, d_{RMSE}))|x \in O\} \tag{12}$$

where $Count(x, O, d_{RMSE})$ and $Count(x, R, d_{RMSE})$ (in equation (3)) calculate the numbers of observed and simulated hotspots within $d_{RMSE}$ range of hotspot $x$. The smaller the difference between $Count(x, O, d_{RMSE})$ and $Count(x, R, d_{RMSE})$, the more similar the hotspot densities of $O$ and $R$ at $x$. From a global perspective, the RMSE of $Com(d_{RMSE})$ can describe the goodness-of-fit. The results are compared using different $d_{RMSE}$ for robustness.

### 5.3. Simulation results

To give a detailed and intuitive simulation result analysis, this sub-section focuses on the specified parameters with $d_{cut} = 1000$ m, $K = 3$, and $\alpha = 1$. More comprehensive results for different parameters are presented in Appendix C.

Figure 14 shows the spatial distribution of the observed local hotspots and an example simulated using different methods. The KNN-based results are very similar to the observed results: they reproduce not only the centers of Hankou (west bank of the river) and Wuchang (east bank of the river), but also the distribution of low-level hotspots that extend outward. However, the hotspots generated by the global contribution-based method are mainly concentrated in the central area of Hankou. This is because Hankou has a significant number of Level 1 hotspots and the global contribution-based process constantly increases the likelihood of generating next lower-level hotspots in this dense hotspots area, forming a Matthew effect (Perc 2014) and lacking the mechanism of outward diffusion distribution. Finally, a completely random result causes the hotspot distribution to be scattered throughout the city area, which differes significantly from the observed results.

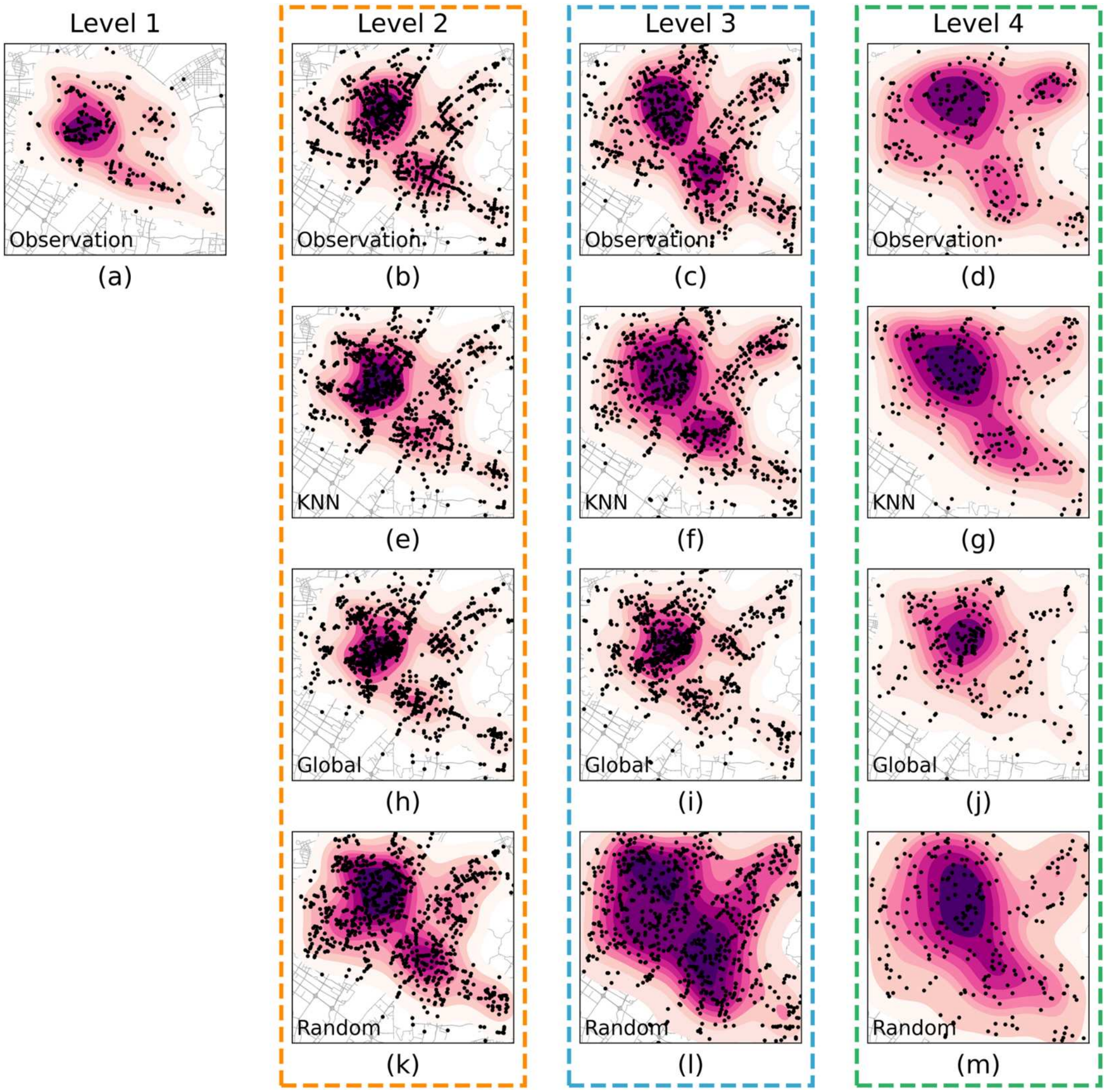

**Figure 14.** Spatial distribution of observed and simulated hotspots with different methods.

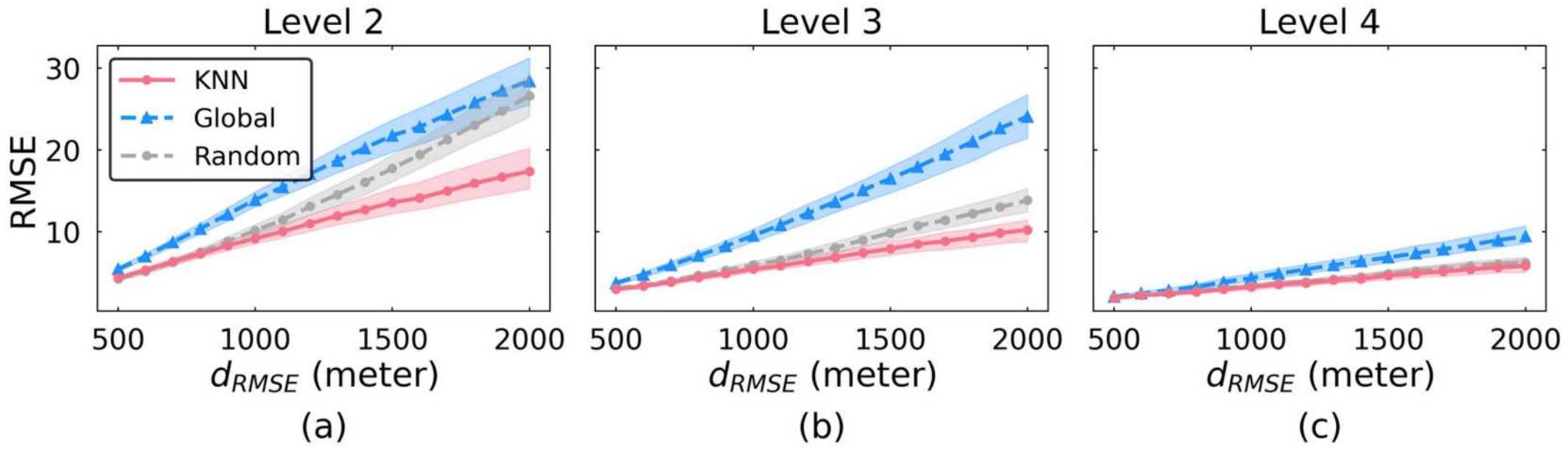

**Figure 15.** RMSE for comparison set between observed hotspots and simulation results with different $d_{RMSE}$ in level 2–4 (a–c). The dashed line is the median value for 100 simulations and the shaded area corresponds to the 10–90% quantile range.

We conducte 100 simulation experiments and calculate the RMSE of the comparison set for each statistical distance $d_{RMSE}$ (Figure 15). Results show that the KNN-based method has lower RMSEs than the global contribution-based and random methods, which indicates the KNN-based method performs better.

## 6. Discussion

### 6.1. Relation between taxi local hotspots and POIs

Individual pick-up and drop-off preferences are largely influenced by POIs in cities. Therefore, analyzing the distribution of taxi local hotspots around POIs helps deepen the understanding of the spatial distribution pattern. Thirteen common POI types, obtained from Amap, Inc. China in 2016, are considered in this study (Table 2).

We do not distinguish the types of POIs and use the LMD method with a 40-m radius parameter to extract the local hotspots of POIs. In addition, the Loubar method is used to classify the levels. This analysis, which is based on local hotspots, provides a relatively fair comparison environment for taxis and POIs. A local hotspot with many POIs is considered a popular area, such as a CBD, which contains many POIs of various types.

Notably, the size or scale is effective to measure the importance of POIs (Wang et al. 2021; Yuan and Bauer 2007). However, our study does not consider the size or scale of POIs, treating all POIs as equally important points. This is due to the lack of information on the area occupied by each POI in our dataset. Although it fails describe the differences of POIs more accurately, previous studies suggested that the analysis based on the POI point density in the region can also yield some valid results in urban-related analysis, including land-use classification (Wu et al. 2021), urban vitality (He et al. 2018; Jiang et al. 2024), and regional spatial interactions (Chen et al. 2024). Therefore, our study remains persuasive to some extent, but future work can focus on incorporating POI size and scale to further refine its conclusions.

**Table 2.** POI data.

| POI data type | Primary information | Number |
|---|---|---|
| Shopping | store, mall, supermarket, … | 88,512 |
| Food | restaurants, bars, coffee shops, … | 49,085 |
| Life service | barber shop, service center, laundromat, … | 47,974 |
| Enterprise | office buildings | 31,618 |
| Education | high school, university, training institution, … | 14,558 |
| Transportation facility | bus station, subway station, … | 14,358 |
| Residential area | apartment, dormitory, mansion, house, … | 11,071 |
| Government | governmental office, patrol station, policy station, … | 8571 |
| Healthcare | hospital, pharmacy, clinic, … | 8530 |
| Recreation | game room, sports center, concert hall, … | 7697 |
| Residential service | hotel | 6115 |
| Public facility | public toilet, evacuation shelter, … | 1971 |
| Tourist attraction | national parks, pavilion, church, … | 1099 |

We focus on the first four levels of POI hotspots, as they account for 1405 of the 1463 hotspots, comprising 99.17% of the total POIs (Appendix D Table D1). Similar to the distribution of taxi pick-up hotspots, there is a significant concentration of POI hotspots on both sides of the river. Notably, the spatial distributions (Figure 16) suggest that although POI hotspots still exhibit an accompanying pattern, no significant inhibiting pattern is observed. In particular, Levels 3 and 4 show a spatial distribution that was quite similar to that of the higher-level hotspots, with no evident spatial diffusion. More quantitative evidence can be found in Appendix D.

To quantify the taxi hotspots around each POI hotspot, we count the number of taxi hotspots within the effective radius of each POI. Considering that the typical walking distance when looking for a taxi is between 100 and 200 m, we set 200 m as the effective radius. Using this approach, a single taxi hotspot may fall within the range of multiple POI hotspots. This is reasonable based on empirical observations, as people often board or alight from taxis in the same location, but may have different destinations.

Figure 17 shows the average number of taxi hotspots within a 200-m radius of each POI local hotspot. Notably, in the study area, the POI hotspots of levels 2–4 had a very similar number of nearby taxi hotspots, averaging more than two each, with Level 2 hotspots having the most, averaging over one. In contrast, Level 1 POI hotspots have an average of 3.5 taxi hotspots, with the difference mainly due to the higher number of Level 1 taxi hotspots.

Therefore, POI hotspots at different levels tend to have multiple taxi hotspots at varying levels around them. This is consistent with the accompanying patterns. We speculate that this phenomenon may stem from people's behavior in selecting pick-up and drop-off locations; people not only frequent popular locations but also choose nearby alternatives (Yuan et al. 2011). For instance, to avoid congestion, passengers may choose relatively spacious areas that are slightly distant from train stations (Qu et al. 2019), or opt for quiet side streets adjacent to busy ones.

In addition, within 200 meters of each POI hotspot, there is an average of only 2–3 taxi hotspots. This indicates a relatively consistent choice of pick-up and drop-off locations within local areas. We suspect this may be one of the reasons for the inhibiting pattern. For instance, level 1 POI hotspots have an average of 1.17 level 1 taxi hotspots, suggesting these areas are commonly used as pick-up locations. However, level 2–4 taxi hotspots around level 1 POI hotspots do not show a significant increase compared to those around level 2–4 POI hotspots (Figure 17b–d). Thus, the statistical relative density exhibits an inhibiting pattern: the relative density of level 2 is not as high as that of level 1. Similarly, the inhibiting pattern is also evident in the relatively smaller presence of level 3 around each POI hotspot than level 2. Across the entire study area, the ratio of level 3 to level 2 is $739/976 \approx 0.757$. Around POI hotspots of the four levels, these ratios are $0.8/1.42 \approx 0.563$, $0.61/1.24 \approx 0.492$, $0.59/1.13 \approx 0.522$, and $0.65/1.11 \approx 0.586$, respectively, which are smaller than the overall ratio. This indicates that the relative density of level 3 around each POI hotspot is lower than that of level 2. A similar pattern exists between level 3 and level 4. Taxi hotspots that are not within 200 m of POI hotspots are relatively evenly distributed throughout the entire urban space (Appendix D, Figure D3).

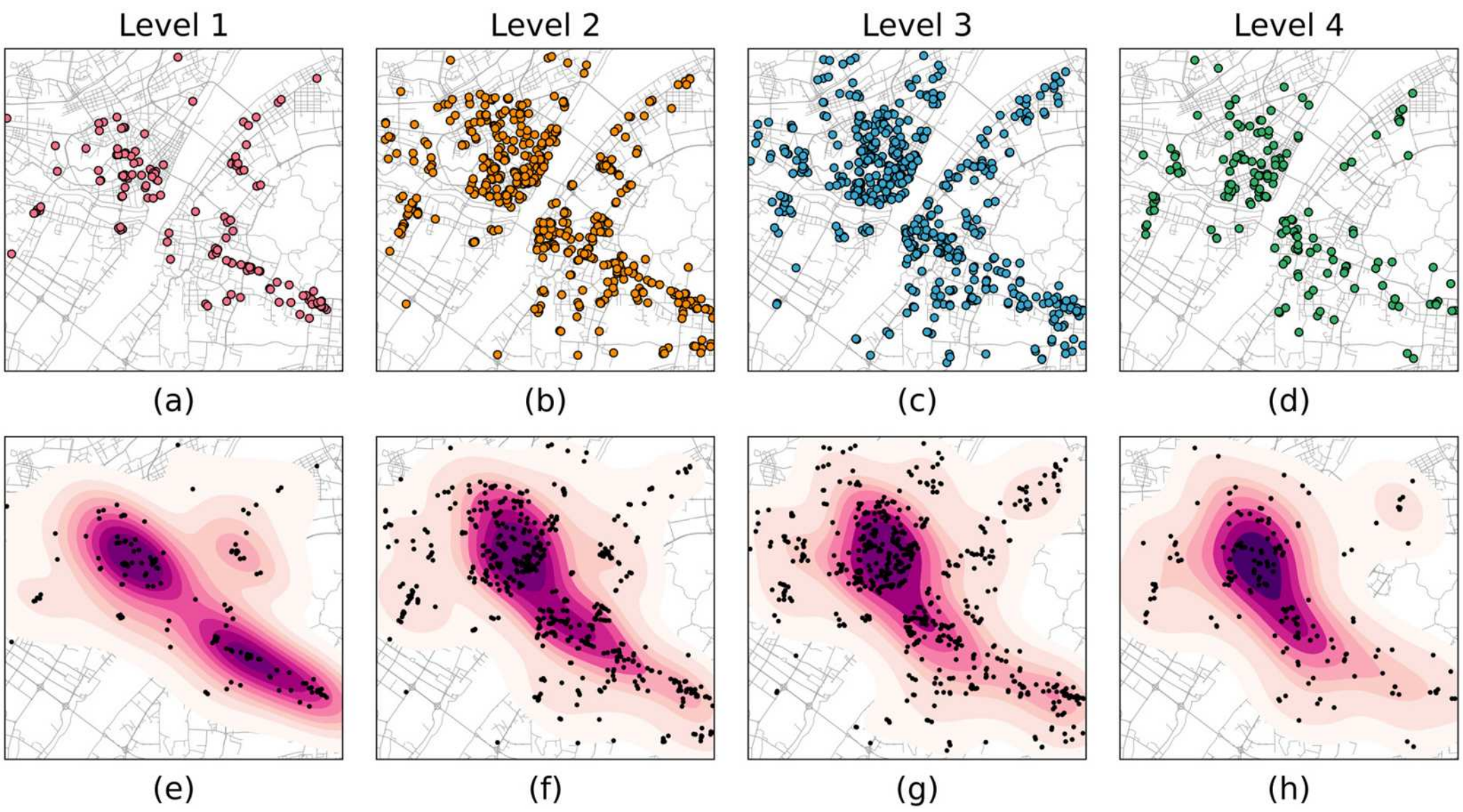

**Figure 16.** Spatial distributions of POI local hotspots (a–d) in Wuhan with the kernel density estimation for visualization purposes (e–h).

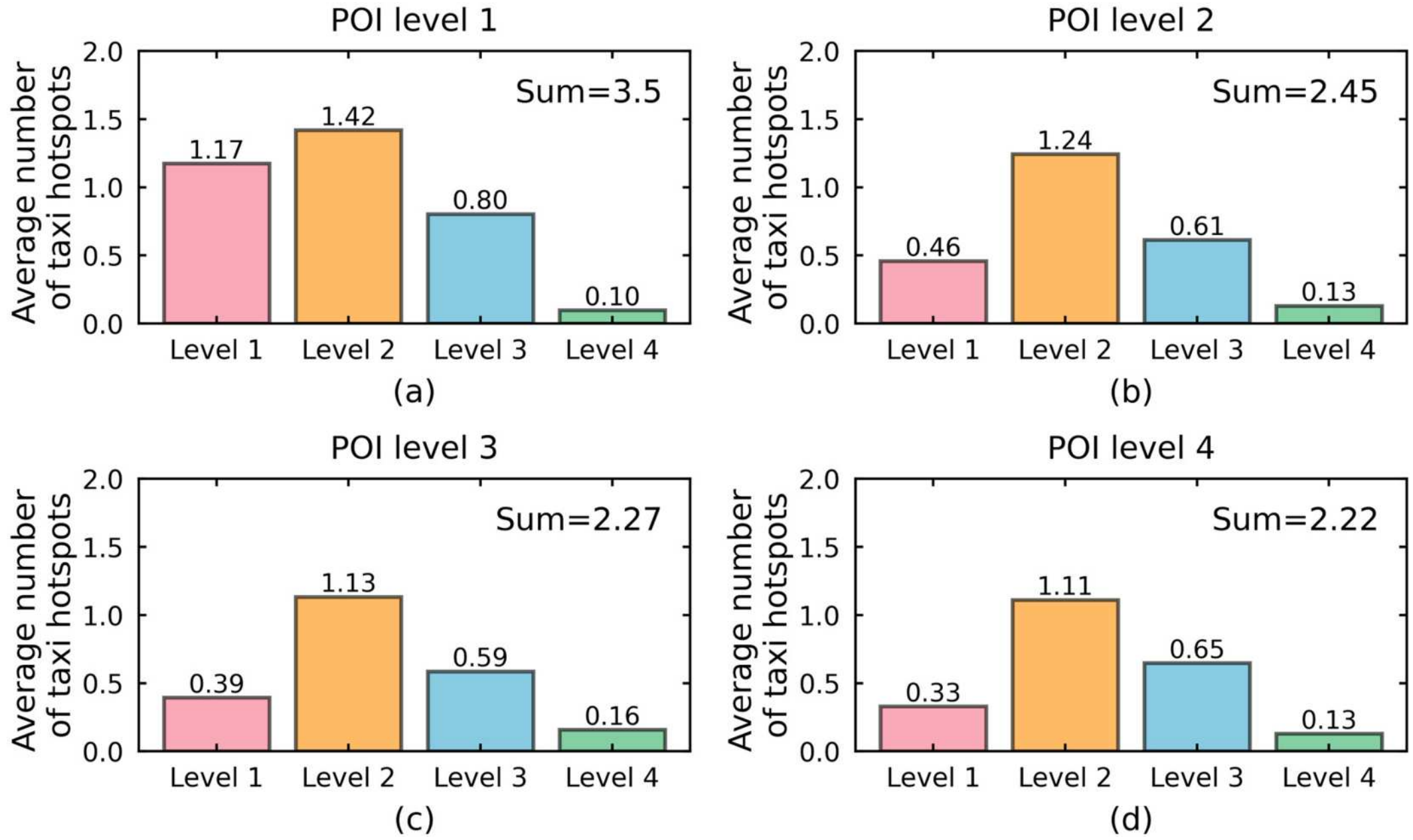

**Figure 17.** Average number of taxi pick-up hotspots within a 200-m radius of each POI local hotspot.

### 6.2. Implications and applications

The local hotspot is an aggregating representation of individuals' activities, characterizing the urban organization structure at a fined scale. The spatial arrangements can be analogized to the mixture of high, medium, and low mountain peaks. There are medium peaks around a high peak (accompanying pattern), but the gap between two high peaks may crowd the space where the medium peaks exist, so the medium peaks have to distribute in the periphery (inhibiting pattern). The medium peaks also affect the low peaks in this way, forming a final mixture arrangement of peaks.

Previous research has indicated that large-scale hotspots of varying intensities exhibit a hierarchical pattern of spatial mixing (Sun and Fan 2021; Zhang et al. 2021). This pattern is similar to the accompanying pattern of local hotspots identified in our study, suggesting that mixed-use patterns of urban areas exist across multiple scales. However, compared with previous studies that mainly examined visually, we strictly use quantitative methods to verify this pattern. In addition, there is an inhibiting pattern in local hotspots, which may be related to the relative consistency of people's local choices of pick-up and drop-off locations. Therefore, both patterns of local hotspots not only deepen our understanding of the micro-scale structure of urban spaces but also provide insights into the spatial selection processes behind individuals' choices for pick-up and drop-off locations.

Our results provide a new perspective on the well-known conjecture of Central Place Theory (CPT) (Mulligan 1984) from the viewpoint of local hotspots. The key idea of CPT argued that each level of the center had its own service scope, and people tend to travel to the nearest center according to the level of demand (Zhong et al. 2017). As such, there was a large spacing between high-level centers, and high-level centers were surrounded by low-level centers, eventually forming a uniform hexagonal network in space. From the perspective of local hotspots, the hierarchical accompanying pattern is similar to the spatial arrangement of centers of different levels advocated by CPT. However, our results are not exactly the same as CPT. The local hotspots here do not involve travel information and have no service scope. It mainly depicts the spatial distribution of small-scale areas that are attractive to people and reflects how the city is specifically used by people. In the popular area, high-level local hotspots are also densely distributed, and there is an inhibitory effect between levels, which is different from the centers' spatial distribution pattern in CPT. Therefore, CPT argues that human mobility forms nested hierarchies, which are similar structures observed in local hotspots. Meanwhile, this novel inhibiting pattern may indicate some additional mechanisms of human mobility in local urban areas.

Our findings of spatial arrangement pattern of local hotspots have several potential benefits. First, based on the network hierarchy formed by local hotspots, urban policymakers can divide city spaces into more refined unit areas, where lower-level hotspots are spatially intermixed with higher-level ones. This refined division allows for more effective arrangement of regional security (Tulumello 2017) and traffic management (Viegas, Miguel Martinez, and Silva 2009), as well as precise control of disease transmission (Tildesley et al. 2010).

Second, the inhibiting pattern suggests urban policymakers can adopt a flat layout when planning the distribution of large POIs to avoid congestion. As shown in Figure 11, the concentration of central business districts (CBDs), large hospitals, and popular tourist attractions leads to a dense distribution of high-level hotspots, with fewer low-level hotspots. By spacing these large POIs farther apart and distributing them across various parts of the city, the clustering of high-level hotspots can be reduced. Based on our analysis of the relationship between local taxi hotspots and POIs, multiple taxi hotspots of varying levels can be established around these large POIs, dispersing foot traffic and alleviating congestion.

Finally, the results can also help develop more sophisticated taxi recommendation services: the system can recommend not only popular hotspots, but also relatively less popular hotspots in the surrounding areas. This dual recommendation can make demand more evenly distributed among different locations, reduce people's waiting time during peak hours, and optimize taxi availability. By diversifying recommendations, the system improves the overall efficiency of the service, enabling it to respond more quickly to real-time demand fluctuations and improve user satisfaction.

## 7. Conclusions and future work

To understand the intrinsic spatial distribution pattern of small-scale hotspots, this study quantitatively explored the spatial arrangement of local hotspots with different levels of popularity. Datasets from Wuhan and Beijing, China, were used for the case studies. The results revealed hierarchical accompanying and inhibiting patterns. Finally, a k-nearest neighbor (KNN)-based model was proposed to reveal the possible underlying mechanism.

There are some limitations to this study and future research. Initially, owing to the limited availability of high-precision location-based data, we only use taxi data from Wuhan and Beijing. In the future, data from online ride hailing, buses, and subways in more cities could be combined with our research for analysis.

Further exploration of the patterns to fully characterize the spatial arrangement of pick-up and drop-off hotspots is crucial. This study quantitatively illustrates the hierarchical accompanying and inhibiting modes and partially characterizes the hotspot arrangement. Other crucial aspects include, but are not limited to, the spatial distribution of hotspots at the same level, the arrangement of higher-level hotspots based on lower-level hotspots, and a more precise depiction of the popularity-distance relationship within the hierarchical structure.

Third, the proposed KNN model cannot simulate the exact boundaries between hotspots, which is a limitation of this study. In future work, the relationship between hotspot boundaries can be studied further to refine the simulation process. In addition, other models can be introduced to reproduce the observed patterns, thereby deepening our understanding of the internal mechanisms of the spatial arrangement of hotspots.

Fourth, the optimal scale of local hotspots in different cities is different, which may affect the comparability of the analysis results between cities. In future work, a general-purpose framework should be developed: the degree of accompanying and suppressing patterns in different cities and local differences can be evaluated while considering the differences in the scale of local hotspots. This will increase the robustness of the method and evaluate the universality and diversity of the discovered patterns.

Finally, urban hotspots are influenced by the urban environment. Studying how urban environments affect the spatial relationship of hotspot popularity is important. In our study, we do not distinguish the importance of different POIs, but only analyzed the number of taxi hotspots around the POI local hotspots. In future work, we can further consider the size and scale of different POIs, and analyze and understand how these differences affect the distribution of taxi hotspots. This is very important to support more accurate and refined urban layout.

## Disclosure statement

No potential conflict of interest was reported by the author(s).

## Funding

This work was supported by the National Natural Science Foundation of China under Grant [number 42201507, 42401560, 42271471, 42271426].

## Data availability statement

The data that support the findings of this study are available from the corresponding author upon reasonable request.

## References

Aslam, Javed, Sejoon Lim, Xinghao Pan, and Daniela Rus. 2012. "City-Scale Traffic Estimation from a Roving Sensor Network." In *Proceedings of the 10th ACM Conference on Embedded Network Sensor Systems*, 141–154. https://doi.org/10.1145/2426656.2426671.

Baddeley, Adrian, Imre Bárány, and Rolf Schneider. 2007. "Spatial Point Processes and Their Applications." In *Stochastic Geometry: Lectures Given at the CIME Summer School Held in Martina Franca, Italy, September 13–18,* 2004, 1–75. https://doi.org/10.1007/978-3-540-38175-4_1.

Bassolas, Aleix, Hugo Barbosa-Filho, Brian Dickinson, Xerxes Dotiwalla, Paul Eastham, Riccardo Gallotti, Gourab Ghoshal, et al. 2019. "Hierarchical Organization of Urban Mobility and Its Connection with City Livability." *Nature Communications* 10 (1): 1–10. https://doi.org/10.1038/s41467-019-12809-y.

Bi, Shuoben, Yuyu Sheng, Wenwu He, Jingjin Fan, and Ruizhuang Xu. 2021a. "Analysis of Travel Hot Spots of Taxi Passengers Based on Community Detection." *Journal of Advanced Transportation* 2021:1–14. https://doi.org/10.1155/2021/6646768.

Bi, Shuoben, Ruizhuang Xu, Aili Liu, Luye Wang, and Lei Wan. 2021b. “Mining Taxi Pick-Up Hotspots Based on Grid Information Entropy Clustering Algorithm.” *Journal of Advanced Transportation* 2021:1–25. https://doi.org/10.1155/2021/5814879.

Chang, Han-wen, Yu-chin Tai, and Jane Yung-Jen Hsu. 2010. “Context-Aware Taxi Demand Hotspots Prediction.” *International Journal of Business Intelligence and Data Mining* 5 (1): 3–18. https://doi.org/10.1504/IJBIDM.2010.030296.

Chen, Xiao-Jian, Ying Wang, Jiayi Xie, Xinyan Zhu, and Jie Shan. 2021. “Urban Hotspots Detection of Taxi Stops with Local Maximum Density.” *Computers, Environment and Urban Systems* 89:101661. https://doi.org/10.1016/j.compenvurbsys.2021.101661.

Chen, Yueping, Yuanqing Wang, Xinyue Zhao, Yongjiu Shi, Hui Zhou, and Yunsheng Li. 2011. “The Fatigue Research of Straight Thread Sleeve Connection for Beijing South Railway Station.” *Applied Mechanics and Materials* 94:1003–1007. https://doi.org/10.4028/www.scientific.net/AMM.94-96.1003.

Chen, Chao, Daqing Zhang, Nan Li, and Zhi-Hua Zhou. 2014. “B-Planner: Planning Bidirectional Night Bus Routes Using Large-Scale Taxi GPS Traces.” *IEEE Transactions on Intelligent Transportation Systems* 15 (4): 1451–1465. https://doi.org/10.1109/TITS.2014.2298892.

Chen, Xiao-Jian, Yuhui Zhao, Chaogui Kang, Xiaoyue Xing, Quanhua Dong, and Yu Liu. 2024. “Characterizing the Temporally Stable Structure of Community Evolution in Intra-Urban Origin-Destination Networks.” *Cities* 150 (July): 105033. https://doi.org/10.1016/j.cities.2024.105033.

Chen, Jinhua, Can Zhao, Kai Zhang, and Zhiheng Li. 2020. “Urban Hotspots Mining Based on Improved FDBSCAN Algorithm.” *Journal of Physics: Conference Series* 1584 (1): 012072. https://doi.org/10.1088/1742-6596/1584/1/012072.

Cui, Haifu, Liang Wu, Sheng Hu, Rujuan Lu, and Shanlin Wang. 2021. “Research on the Driving Forces of Urban Hot Spots Based on Exploratory Analysis and Binary Logistic Regression Model.” *Transactions in GIS* 25 (3): 1522–1541. https://doi.org/10.1111/tgis.12739.

Elhorst, J Paul, Donald J Lacombe, and Gianfranco Piras. 2012. “On Model Specification and Parameter Space Definitions in Higher Order Spatial Econometric Models.” *Regional Science and Urban Economics* 42 (1-2): 211–220. https://doi.org/10.1016/j.regsciurbeco.2011.09.003.

Faghih-Imani, Ahmadreza, Sabreena Anowar, Eric J Miller, and Naveen Eluru. 2017. “Hail a Cab or Ride a Bike? A Travel Time Comparison of Taxi and Bicycle-Sharing Systems in New York City.” *Transportation Research Part A: Policy and Practice* 101:11–21. https://doi.org/10.1016/j.tra.2017.05.006.

He, Liang, Suozhong Chen, Ying Liang, Manqin Hou, and Junyi Chen. 2020. “Infilling the Missing Values of Groundwater Level Using Time and Space Series: Case of Nantong City, East Coast of China.” *Earth Science Informatics* 13 (4): 1445–1459. https://doi.org/10.1007/s12145-020-00489-y.

He, Qingsong, Weishan He, Yan Song, Jiayu Wu, Chaohui Yin, and Yanchuan Mou. 2018. “The Impact of Urban Growth Patterns on Urban Vitality in Newly Built-up Areas Based on an Association Rules Analysis Using Geographical “Big Data”.” *Land Use Policy* 78:726–738. https://doi.org/10.1016/j.landusepol.2018.07.020.

Hulley, Glynn, Sarah Shivers, Erin Wetherley, and Robert Cudd. 2019. “New ECOSTRESS and MODIS Land Surface Temperature Data Reveal Fine-Scale Heat Vulnerability in Cities: A Case Study for Los Angeles County, California.” *Remote Sensing* 11 (18): 2136. https://doi.org/10.3390/rs11182136.

Illian, Janine, Antti Penttinen, Helga Stoyan, and Dietrich Stoyan. 2008. *Statistical Analysis and Modelling of Spatial Point Patterns*. Chichester, UK: Wiley.

Jiang, Bin. 2013. “Head/Tail Breaks: A New Classification Scheme for Data with a Heavy-Tailed Distribution.” *The Professional Geographer* 65 (3): 482–494. https://doi.org/10.1080/00330124.2012.700499.

Jiang, Yanxiao, Zhou Huang, Xiao Zhou, and Xiaojian Chen. 2024. “Evaluating the Impact of Urban Morphology on Urban Vitality: An Exploratory Study Using Big Geo-Data.” *International Journal of Digital Earth* 17 (1): 2327571. https://doi.org/10.1080/17538947.2024.2327571.

Kumar, Dheeraj, Huayu Wu, Yu Lu, Shonali Krishnaswamy, and Marimuthu Palaniswami. 2016. “Understanding Urban Mobility via Taxi Trip Clustering.” In *2016 17th IEEE International Conference on Mobile Data Management (MDM)*, 1, 318–324. https://doi.org/10.1109/MDM.2016.54.

Li, Ke, Haitao Wei, Xiaohui He, and Zhihui Tian. 2022. “Relational POI Recommendation Model Combined with Geographic Information.” *PLoS One* 17 (4): e0266340. https://doi.org/10.1371/journal.pone.0266340.

Liu, Yunzhe, Alex Singleton, Daniel Arribas-Bel, and Meixu Chen. 2021. “Identifying and Understanding Road-Constrained Areas of Interest (AOIs) Through Spatiotemporal Taxi GPS Data: A Case Study in New York City.” *Computers, Environment and Urban Systems* 86:101592. https://doi.org/10.1016/j.compenvurbsys.2020.101592.

Louail, Thomas, Maxime Lenormand, Oliva G. Cantu Ros, Miguel Picornell, Ricardo Herranz, Enrique Frias-Martinez, José J. Ramasco, and Marc Barthelemy. 2014. “From Mobile Phone Data to the Spatial Structure of Cities.” *Scientific Reports* 4 (1): 5276. https://doi.org/10.1038/srep05276.

May, Michael, Dirk Hecker, Christine Körner, Simon Scheider, and Daniel Schulz. 2008. “A Vector-Geometry Based Spatial Knn-Algorithm for Traffic Frequency Predictions.” In *2008 IEEE International Conference on Data Mining Workshops*, 442–447. IEEE. https://doi.org/10.1109/ICDMW.2008.35.

Miller, Harvey J, Somayeh Dodge, Jennifer Miller, and Gil Bohrer. 2019. “Towards an Integrated Science of Movement: Converging Research on Animal Movement Ecology and Human Mobility Science.” *International Journal of Geographical Information Science* 33 (5): 855–876. https://doi.org/10.1080/13658816.2018.1564317.

Mulligan, Gordon F. 1984. “Agglomeration and Central Place Theory: A Review of the Literature.” *International Regional Science Review* 9 (1): 1–42. https://doi.org/10.1177/016001768400900101.

Nong, Yun, Suhong Zhou, Lin Liu, Qiuping Li, Yinong Peng, and Xinhua Hao. 2019. “Structural Cities: Delimiting Retailing Center Boundaries and Their Hierarchical Characteristics in Urban China Based on GPS-Enabled Taxi Data.” *Journal of Planning Education and Research* 39 (3): 345–359. https://doi.org/10.1177/0739456X17741964.

Palaniswami, Marimuthu, Aravinda S Rao, Dheeraj Kumar, Punit Rathore, and Sutharshan Rajasegarar. 2020. “The Role of Visual Assessment of Clusters for Big Data Analysis: From Real-World Internet of Things.” *IEEE Systems, Man, and Cybernetics Magazine* 6 (4): 45–53. https://doi.org/10.1109/MSMC.2019.2961160.

Pan, Gang, Guande Qi, Zhaohui Wu, Daqing Zhang, and Shijian Li. 2012. “Land-Use Classification Using Taxi GPS Traces.” *IEEE Transactions on Intelligent Transportation Systems* 14 (1): 113–123. https://doi.org/10.1109/Tits.2012.2209201.

Perc, Matjaž. 2014. “The Matthew Effect in Empirical Data.” *Journal of The Royal Society Interface* 11 (98): 20140378. https://doi.org/10.1098/rsif.2014.0378.

Qu, Zhaowei, Xin Wang, Xianmin Song, Zhaotian Pan, and Haitao Li. 2019. “Location Optimization for Urban Taxi Stands Based on Taxi GPS Trajectory Big Data.” *IEEE Access* 7:62273–62283. https://doi.org/10.1109/ACCESS.2019.2916342.

Sila-Nowicka, Katarzyna, Jan Vandrol, Taylor Oshan, Jed A Long, Urška Demšar, and A. Stewart Fotheringham. 2016. “Analysis of Human Mobility Patterns from GPS Trajectories and Contextual Information.” *International Journal of Geographical Information Science* 30 (5): 881–906. https://doi.org/10.1080/13658816.2015.1100731.

Sun, Mengqi, and Hongchao Fan. 2021. “Detecting and Analyzing Urban Centers Based on the Localized Contour Tree Method Using Taxi Trajectory Data: A Case Study of Shanghai.” *ISPRS International Journal of Geo-Information* 10 (4): 220. https://doi.org/10.3390/ijgi10040220.

Tildesley, Michael J., Thomas A. House, Mark C. Bruhn, Ross J. Curry, Maggie O’Neil, Justine L. E. Allpress, Gary Smith, and Matt J. Keeling. 2010. “Impact of Spatial Clustering on Disease Transmission and Optimal Control.” *Proceedings of the National Academy of Sciences* 107 (3): 1041–1046. https://doi.org/10.1073/pnas.0909047107

Tulumello, Simone. 2017. “Toward a Critical Understanding of Urban Security Within the Institutional Practice of Urban Planning: The Case of the Lisbon Metropolitan Area.” *Journal of Planning Education and Research* 37 (4): 397–410. https://doi.org/10.1177/0739456X16664786

Um, Sun-Bi, and Jung-Sup Um. 2015. “Metropolitan Urban Hotspots of Chronic Sleep Deprivation: Evidence from a Community Health Survey in Gyeongbuk Province, South Korea.” *Geospatial Health* 10:2. https://doi.org/10.4081gh.2015.382.

Viegas, José Manuel, L. Miguel Martinez, and Elisabete A. Silva. 2009. “Effects of the Modifiable Areal Unit Problem on the Delineation of Traffic Analysis Zones.” *Environment and Planning B: Planning and Design* 36 (4): 625–643. https://doi.org/10.1068/b34033

Wang, Ziyi, Debin Ma, Dongqi Sun, and Jingxiang Zhang. 2021. “Identification and Analysis of Urban Functional Area in Hangzhou Based on OSM and POI Data.” *PLoS One* 16 (5): e0251988. https://doi.org/10.1371/journal.pone.0251988.

Wu, Rong, Jieyu Wang, Dachuan Zhang, and Shaojian Wang. 2021. “Identifying Different Types of Urban Land Use Dynamics Using Point-of-Interest (POI) and Random Forest Algorithm: The Case of Huizhou, China.” *Cities* 114:103202. https://doi.org/10.1016/j.cities.2021.103202.

Yuan, Fei, and Marvin E Bauer. 2007. “Comparison of Impervious Surface Area and Normalized Difference Vegetation Index as Indicators of Surface Urban Heat Island Effects in Landsat Imagery.” *Remote Sensing of Environment* 106 (3): 375–386. https://doi.org/10.1016/j.rse.2006.09.003.

Yuan, Yihong, and Martin Raubal. 2014. “Measuring Similarity of Mobile Phone User Trajectories–a Spatio-Temporal Edit Distance Method.” *International Journal of Geographical Information Science* 28 (3): 496–520. https://doi.org/10.1080/13658816.2013.854369.

Yuan, Jing, Yu Zheng, Liuhang Zhang, Xing Xie, and Guangzhong Sun. 2011. “Where to Find My next Passenger.” In *Proceedings of the 13th International Conference on Ubiquitous Computing*, 109–118. https://doi.org/10.1145/2030112.2030128.

Zhang, Yan, Xiang Zheng, Min Chen, Yingbing Li, Yingxue Yan, and Peiying Wang. 2021. “Urban Fine-Grained Spatial Structure Detection Based on a New Traffic Flow Interaction Analysis Framework.” *ISPRS International Journal of Geo-Information* 10 (4): 227. https://doi.org/10.3390/ijgi10040227.

Zheng, Xudong, Xiao Liang, and Ke Xu. 2012. “Where to Wait for a Taxi?” In *Proceedings of the ACM SIGKDD International Workshop on Urban Computing*, 149–156.

Zhong, Chen, Markus Schläpfer, Stefan Müller Arisona, Michael Batty, Carlo Ratti, and Gerhard Schmitt. 2017. “Revealing Centrality in the Spatial Structure of Cities from Human Activity Patterns.” *Urban Studies* 54 (2): 437–455. https://doi.org/10.1177/0042098015601599.

Zhou, Tong, Xintao Liu, Zhen Qian, Haoxuan Chen, and Fei Tao. 2019. "Automatic Identification of the Social Functions of Areas of Interest (AOIs) Using the Standard Hour-Day-Spectrum Approach." *ISPRS International Journal of Geo-Information* 9 (1): 7. https://doi.org/10.3390/ijgi9010007.

# Appendices

## Appendix A. Details of simulation

Formally, the simulation problem is stated below:

(1) (Precondition) The level of hotspots is $1 \leq i \leq m$, the radius setting of a hotspot is $x_{radius}$ m, and the cut-off range is $d_{cut}$ m.
(2) (Notation) $O_i$, $B_i$, and $S_i$ are the set of observed, background, and simulated hotspots, where

$$B_1 = O_1 \tag{A1}$$

$$B_i = \left\{ x \in O_i \middle| \max_{y \in \cup_{1 \leq k \leq i} B_{k-1}} d(x, y) > d_{cut} \right\}, \ 2 \leq i \leq m \tag{A2}$$

For simplicity, we represent $H_i = S_i \cup B_i$.

(3) (Purpose) Generate $S_i$ by $H_{i-1} = S_{i-1} \cup B_{i-1}$ for $2 \leq i \leq m$. $|S_i|$ is determined in advance by $|S_i| = |O_i| - |B_i|$, $i \geq 2$, since this study aims to reveal the possible mechanism through simulation, rather than the number. For $i = 2$, it only uses $B_1$ to generate $S_2$ since $S_1 = \emptyset$.

For the simulation, we select grids overlapping with the road network as the original candidate grid set $G_0$. We describe how to generate level 2 using level 1 on the $G_0$, and the subsequent generation procedure is similar.

(1) Candidate grid set update. Exclude the grids of background hotspots to form a new candidate grid set $G_1$.
(2) Calculate the attraction of a grid based on higher-level hotspots, i.e. $Attr(x|B_1)$.
(3) Randomly pick hotspots. An iterative process is conducted to pick hotspots. For the first pick, a grid $x_1 \in G_1$ is randomly selected according to the probability distribution of $G_1$, where

$$p(x|B_1) = \frac{Attr(x|B_1)}{\sum_{x \in G_1} Attr(x|B_1)} \tag{A3}$$

Then $x_1$ and $x_{radius}$m radius square of $x_1$ are excluded from $G_1$ to form the alternative set for the next iteration. This pick is repeated until all $|O_2| - |B_2|$ local hotspots are selected and the remaining grids form $G_2$.

Similarly, we next use $H_2 = B_2 \cup S_2$ to generate $S_3$ at the base of $G_2$. As such, this procedure generates hotspots level by level.

## Appendix B. Identification of local hotspots and level classification

The study area of Beijing is within the 5th ring road (Figure B1). Information on the study area, time and data volume is recorded in Table B1, where data in 30 March 2015 is removed due to the large volume of missing data caused by equipment failure of the taxi company platform. The basic unit of research is also 10 m.

**Table B1.** Information of taxi data set in Beijing, China.

| | |
|---|---|
| Area | [116.192°E, 116.545°E]×[39.754°N, 40.032°N] (Around 30.3 km×30.6 km) |
| Time | 1 March 2015 to 31 March 2015 |
| Number of pick-up and drop-off stop pairs | 10,153,740 |

The radius of Wuhan's pick-up and drop-off local hotspot is 40 and 30 m by 'elbow point' method proposed by previous work. It further recommends that, as a scale reflecting the movement of people in the city, the radius of the pick-up and drop-off local hotspot should be uniformly selected with a larger value, so as to cover more information, if the difference between the two is not large. As such, we finally set 40 m as the radius for pick-up and drop-off local hotspot. Similarly, the radius of Beijing's pick-up and drop-off local hotspot are 50 and 40 m, respectively, and we finally set 50 m as the uniform scale.

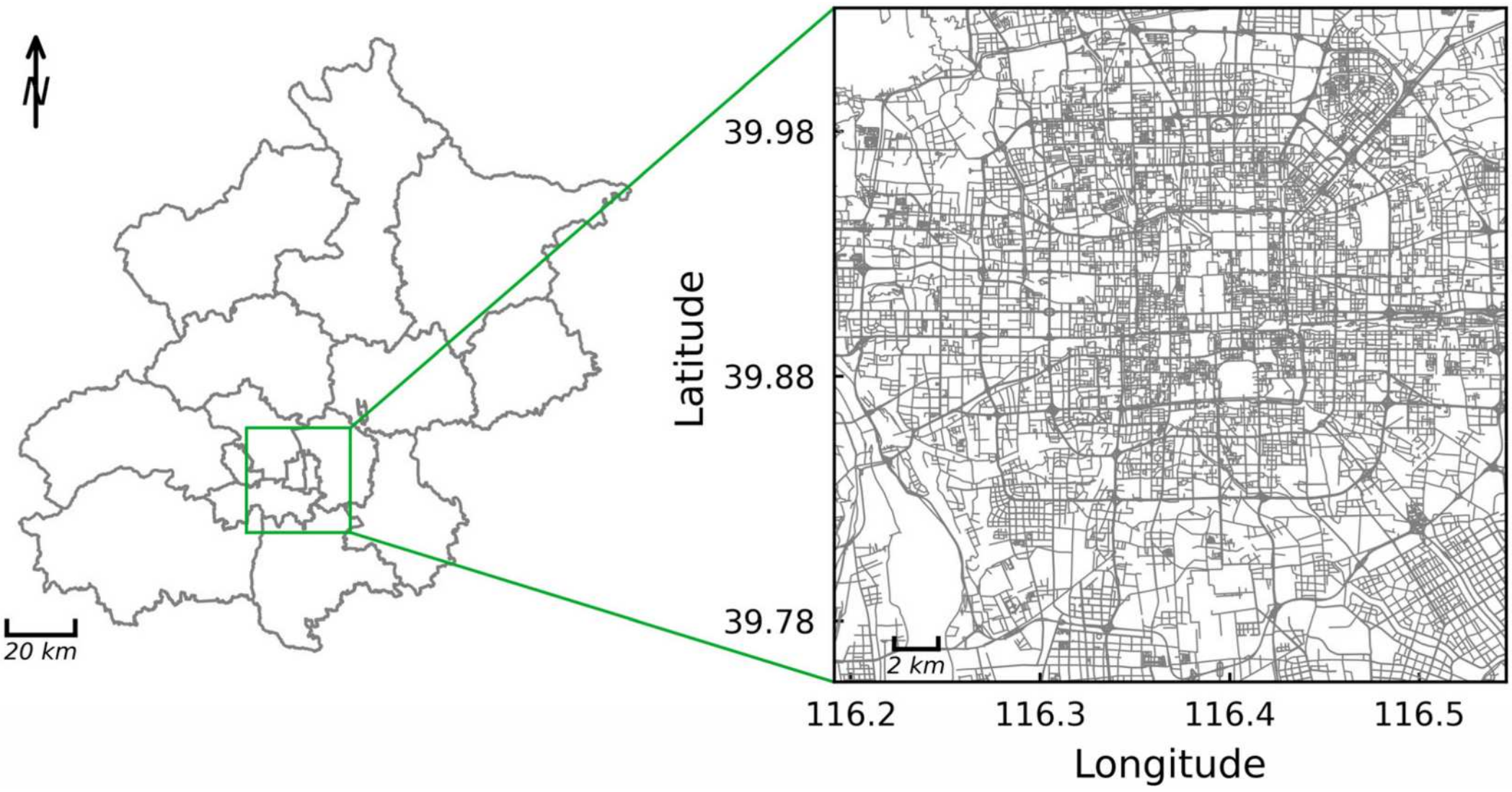

**Figure B1.** The study area of Beijing, China (within the 5-th ring road) and its road network distribution.

In all dataset, the statistical results of the classification results show that most of local hotspots concentrate on the first four levels (Tables B2–B4). Therefore, we only focus on the first four levels in the following results. Figure B2 exhibit their spatial distributions.

**Table B2.** The statistical results of drop-off local hotspots in different levels classified by Loubar method (Wuhan).

| Level | Number of hotspots | Range value of fractions of total stops | Range value of stops (median) |
|---|---|---|---|
| 1 | 84 | [0, 14.57%] | 217,710–17,945 (25,026) |
| 2 | 876 | [14.57%, 57.73%] | 17,930–5864 (8077) |
| 3 | 1326 | [57.73%, 87.99%] | 5860–3023 (4094) |
| 4 | 730 | [87.99%, 98.39%] | 3022–2211 (2609) |
| 5 | 80 | [98.39%,99.83%] | 2209–1576 (2018) |
| 6 | 73 | [99.83%,99.98%] | 1551–1266 (1408) |
| 7 | 3 | [99.98%,99.99%] | 1229–955 (1202) |
| 8 | 1 | [99.99%,100%] | 371 |

**Table B3.** The statistical results of pick-up local hotspots in different levels classified by Loubar method (Beijing).

| Level | Number of hotspots | Range value of fractions of total stops | Range value of stops (median) |
|---|---|---|---|
| 1 | 135 | [0, 16.32%] | 27,712–2733 (3377) |
| 2 | 1149 | [16.32%, 64.88%] | 2729–1115 (1469) |
| 3 | 1180 | [64.88%, 93.30%] | 1114–730 (893) |
| 4 | 344 | [93.30%, 99.11%] | 729–476 (647) |
| 5 | 70 | [99.11%,99.89%] | 474–313 (421) |
| 6 | 13 | [99.89%,99.98%] | 312–185 (299) |
| 7 | 2 | [99.98%,99.99%] | 154–151 (152.5) |
| 8 | 1 | [99.99%,100%] | 139 (139) |

**Table B4.** The statistical results of drop-off local hotspots in different levels classified by Loubar method (Beijing).

| Level | Number of hotspots | Range value of fractions of total stops | Range value of stops (median) |
|---|---|---|---|
| 1 | 179 | [0, 17.15%] | 18,843–2157 (2645) |
| 2 | 1303 | [17.15%, 66.11%] | 2153–879 (1192) |
| 3 | 1306 | [66.11%, 93.19%] | 879–572 (691) |
| 4 | 404 | [93.19%, 99.17%] | 571–375 (505) |
| 5 | 79 | [99.17%,99.89%] | 375–269 (334) |
| 6 | 14 | [99.89%,99.98%] | 260–180 (249.5) |
| 7 | 2 | [99.98%,99.99%] | 172–104 (138) |
| 8 | 1 | [99.99%,100%] | 98 (98) |

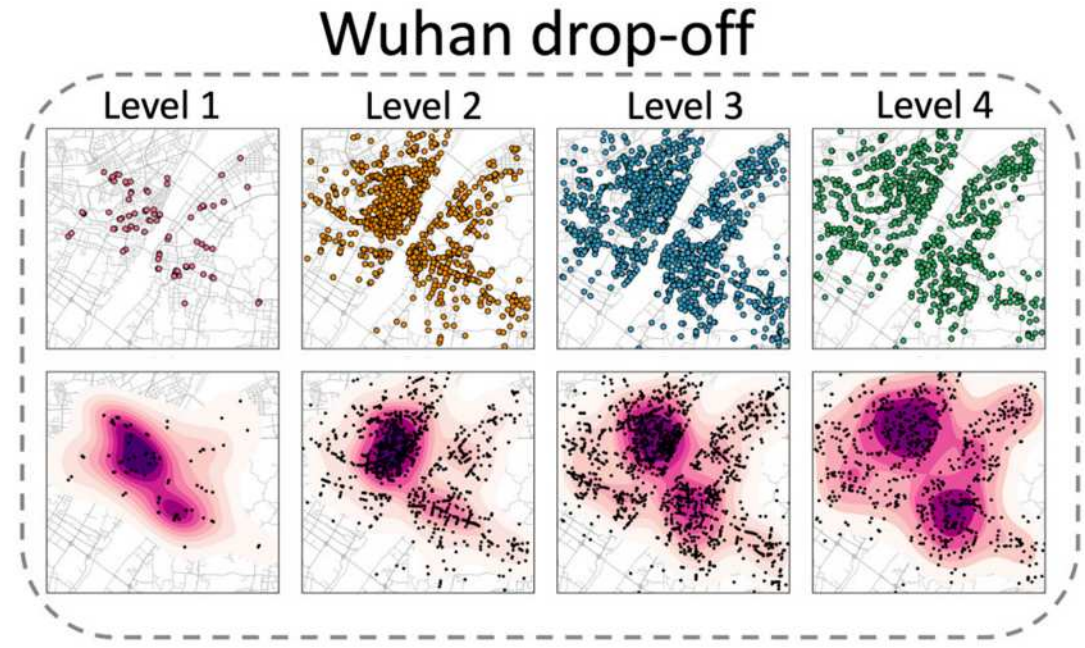

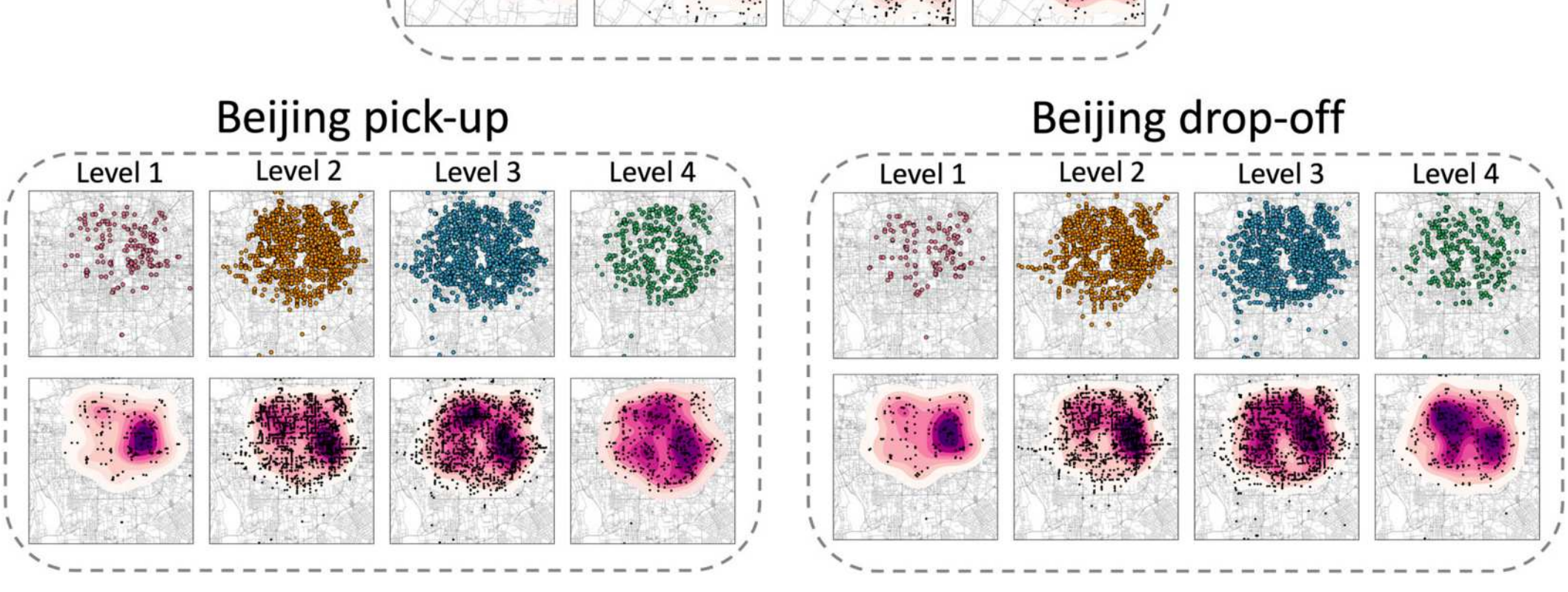

**Figure B2.** The spatial distribution of local hotspots in Wuhan and Beijing.

## *Appendix C. Empirical and simulation results*

Figure C1 Illustrates the indicators to describe accompanying and inhibiting patters. The results are similar as the ones in the main body, supporting the existences of both patterns in these three cases.

Figures C2–C5 show the RMSEs under different parameters. Comparing with our KNN-based approach, global contribution-based and completed random methods cannot perform coincidentally well at all levels. However, in a few cases, the simulation results of the KNN-based method are not better than the global contribution-based method, especially for the simulation results for pick-up local hotspots in Beijing. This may be due to the small number of Level 1 hotspots. However, in general, the KNN-based local mechanism is more suitable for expressing the hierarchical spatial distribution relationship between hotspots of different levels than the global contribution and random cases.

## *Appendix D. Detail information of POI local hotspots*

**Table D1.** The statistical results of different levels of POI local hotspots classified by Loubar method.

| Level | Number of hotspots | Range value of fractions of total stops | Range value of stops (median) |
|---|---|---|---|
| 1 | 143 | [0,31.95%] | 628–121 (166) |
| 2 | 507 | [31.95%,69.94%] | 121–47 (61) |
| 3 | 576 | [69.94%,94.42%] | 46–28 (39) |
| 4 | 179 | [94.42%,99.17%] | 28–17 (24) |
| 5 | 43 | [99.17%,99.87%] | 16–11 (15) |
| 6 | 10 | [99.87%,99.97%] | 11–7 (9) |
| 7 | 4 | [99.97%,99.99%] | 6–5 (5.5) |
| 8 | 1 | [99.99%,100%] | 3 |

Figure D1 confirms the presence of the accompanying pattern. The inhibition pattern is not significant for level 3 compared to level 2 – at most scales ($d_{count}$), the ratio of same Level to next level $Den_{nor}$ remains largely consistent (see Figure D2b). Similarly, Level 4 is not significantly different from Level 3 (see Figure D3b).

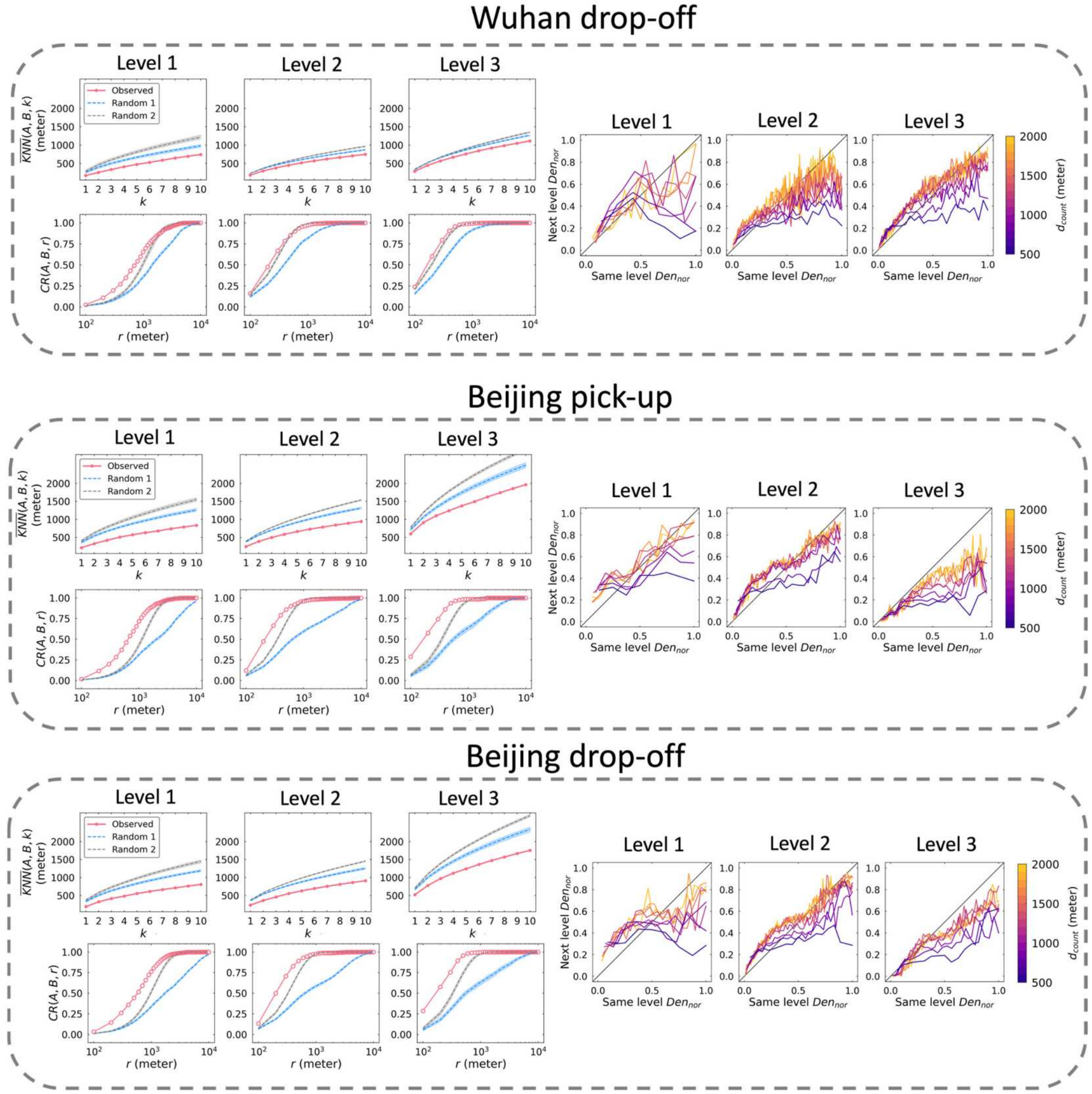

**Figure C1.** The evidences of the accompanying and inhibiting patterns in Wuhan and Beijing.

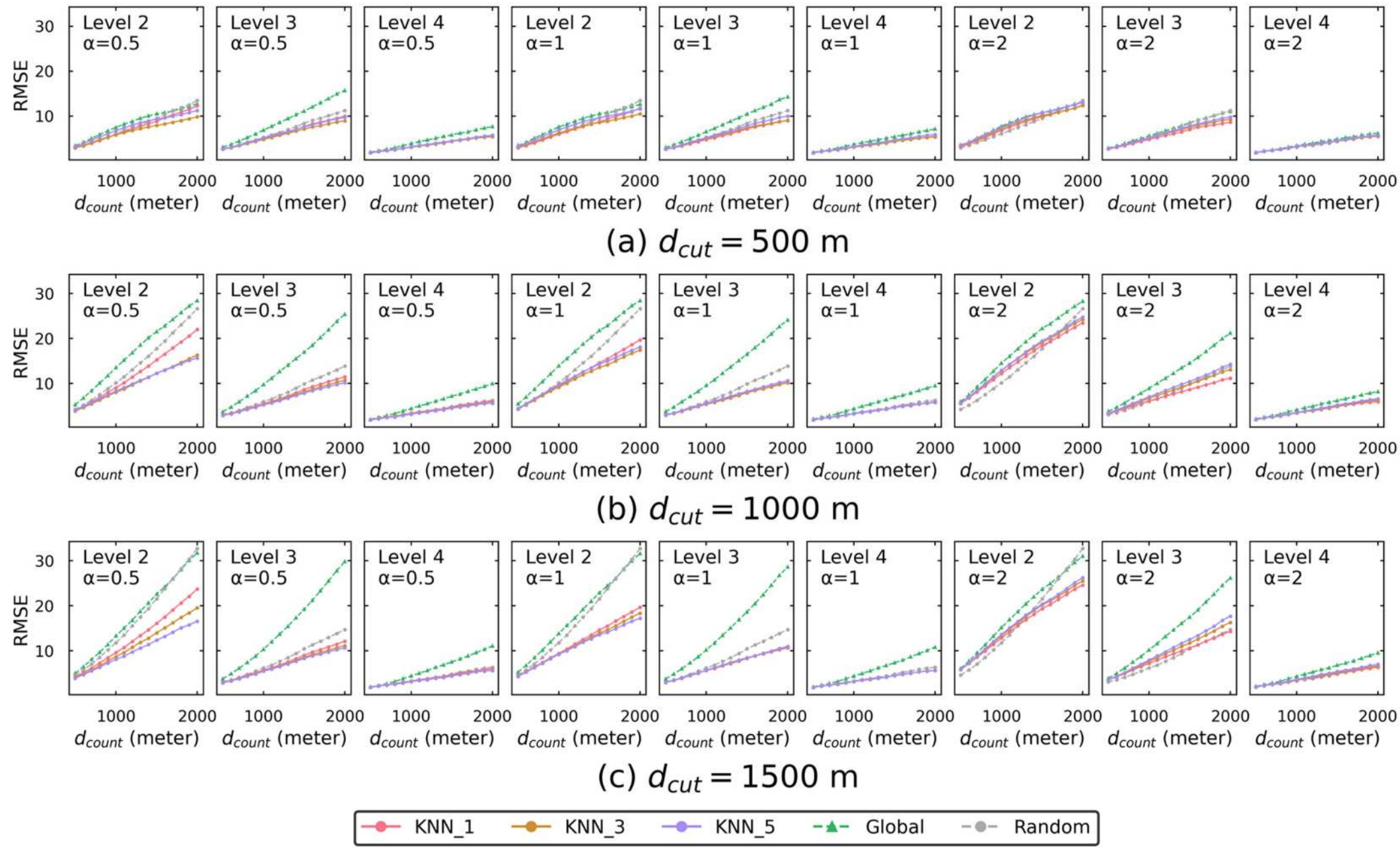

**Figure C2.** Comparison of simulation results of different methods: the pick-up situation in Wuhan.

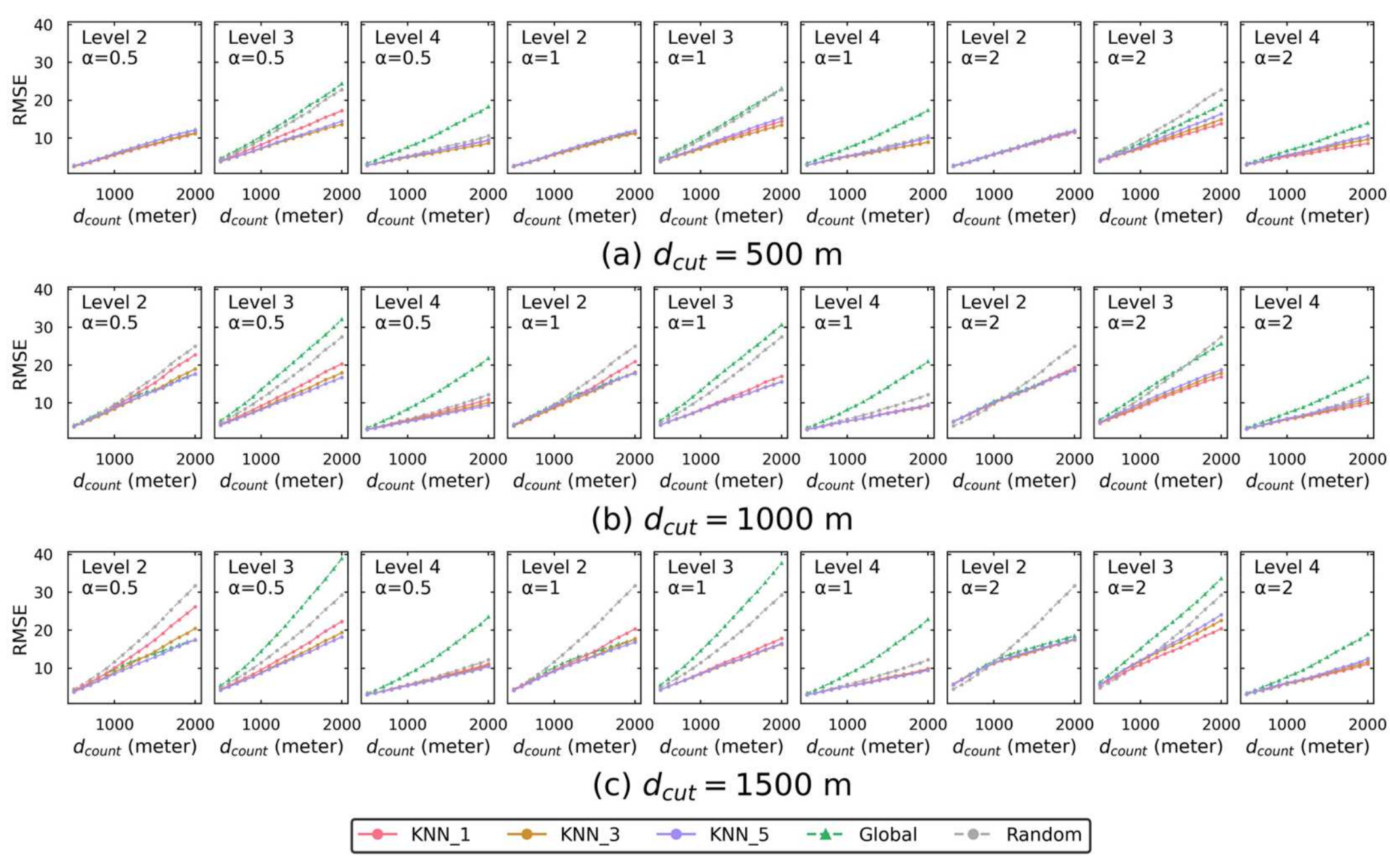

**Figure C3.** Comparison of simulation results of different methods: the drop-off situation in Wuhan.

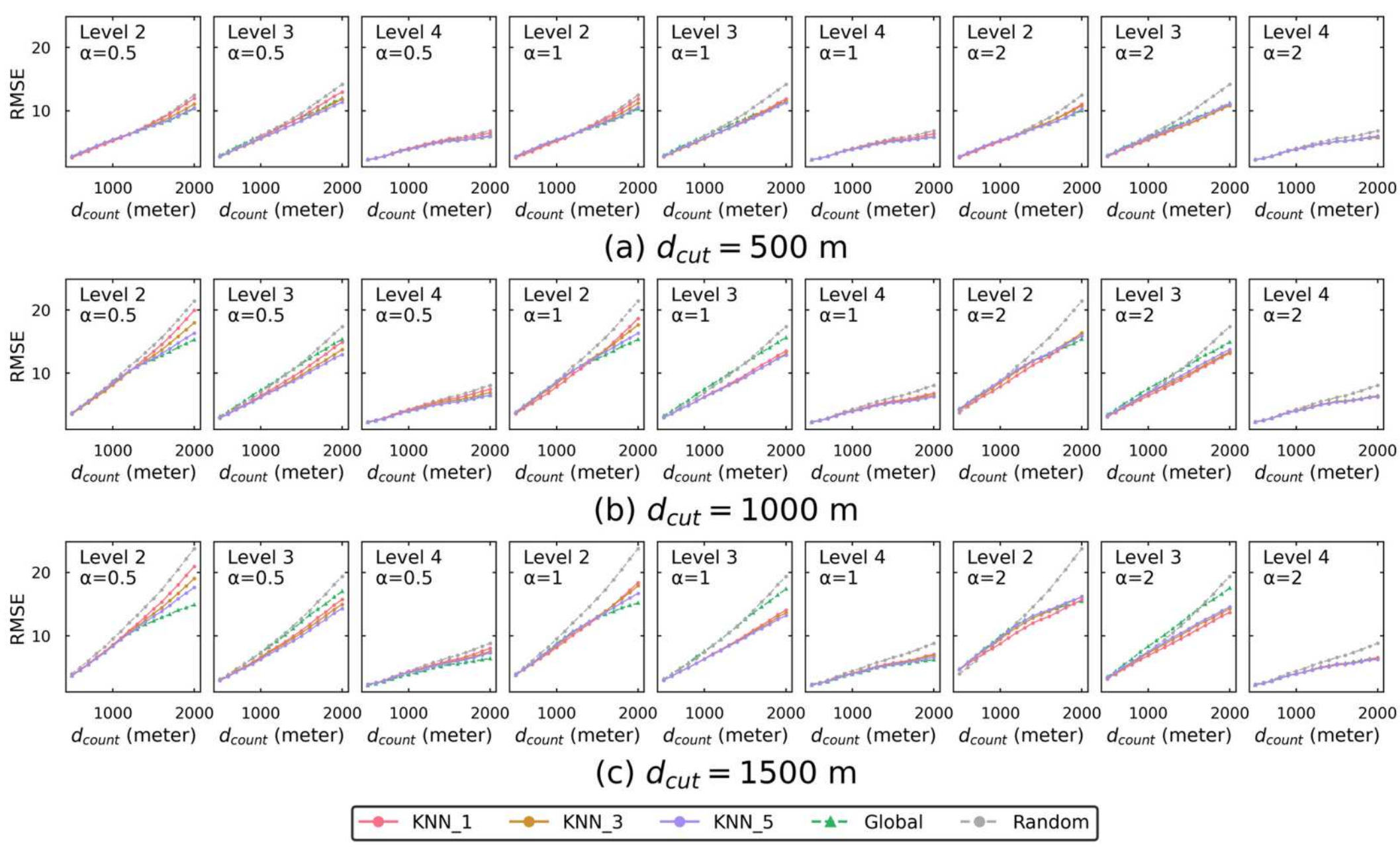

**Figure C4.** Comparison of simulation results of different methods: the pick-up situation in Beijing.

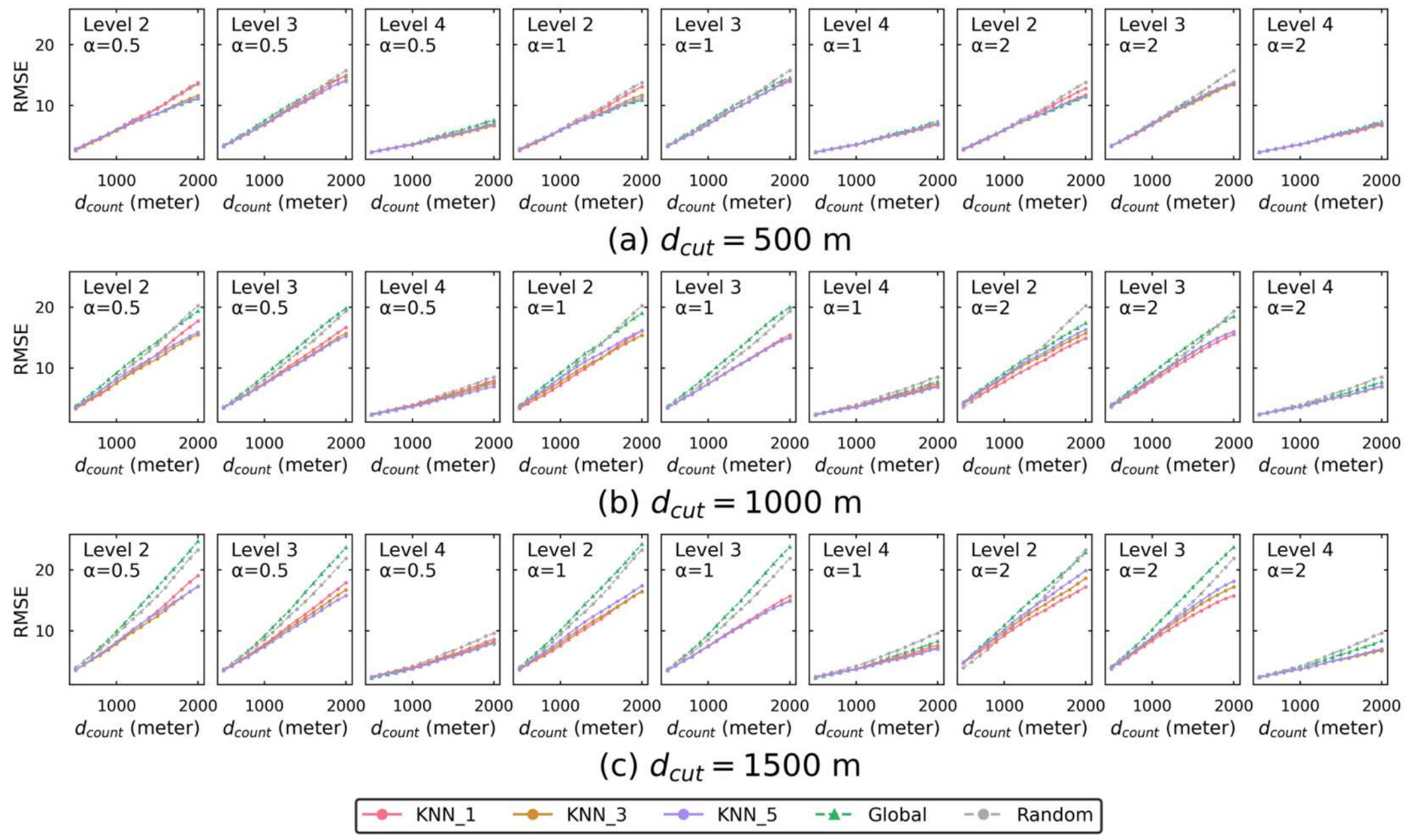

**Figure C5.** Comparison of simulation results of different methods: the drop-off situation in Beijing.

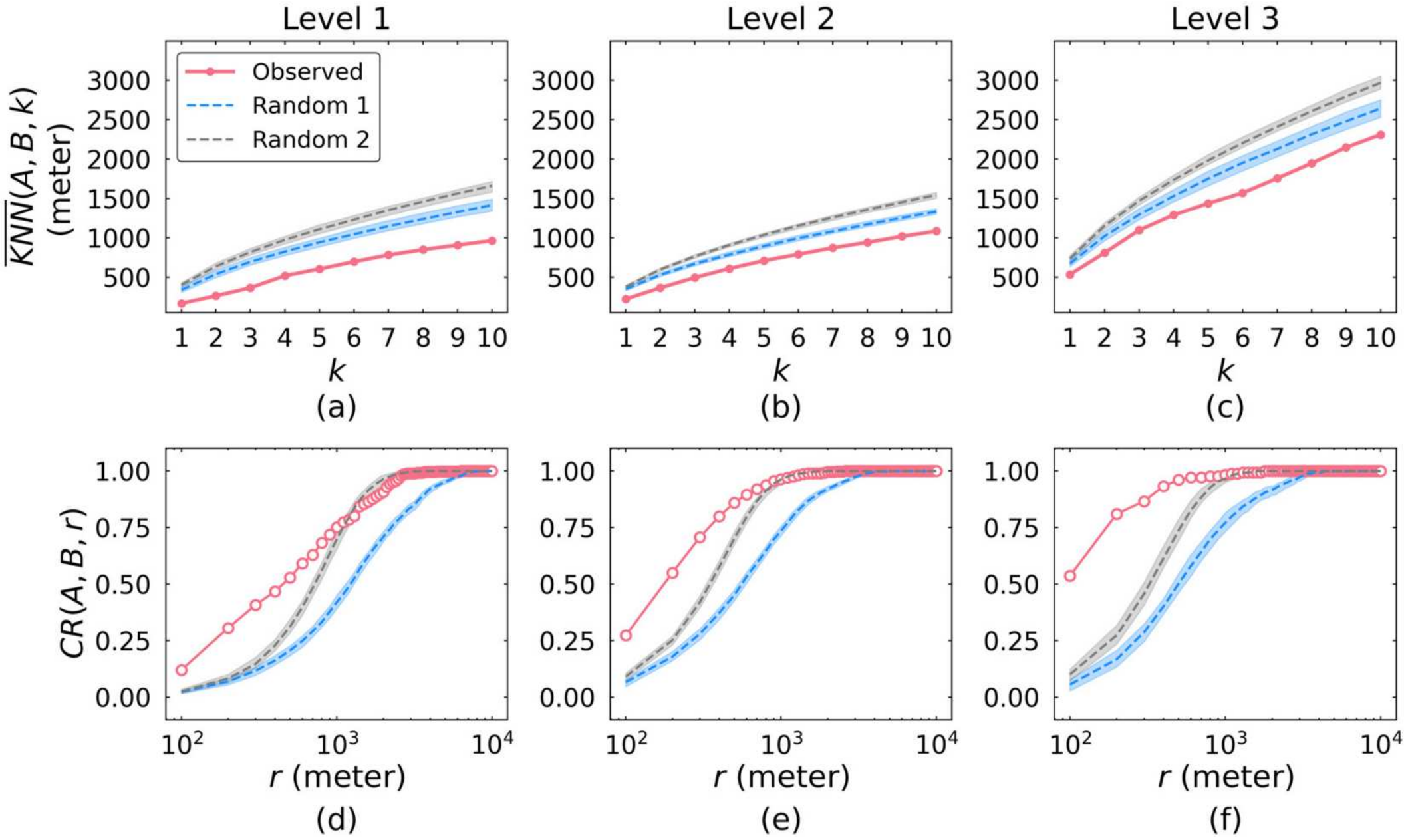

**Figure D1.** The $\overline{KNN}(A, B, k)$ and $CR(A, B, r)$ for POI local hotspots in Wuhan.

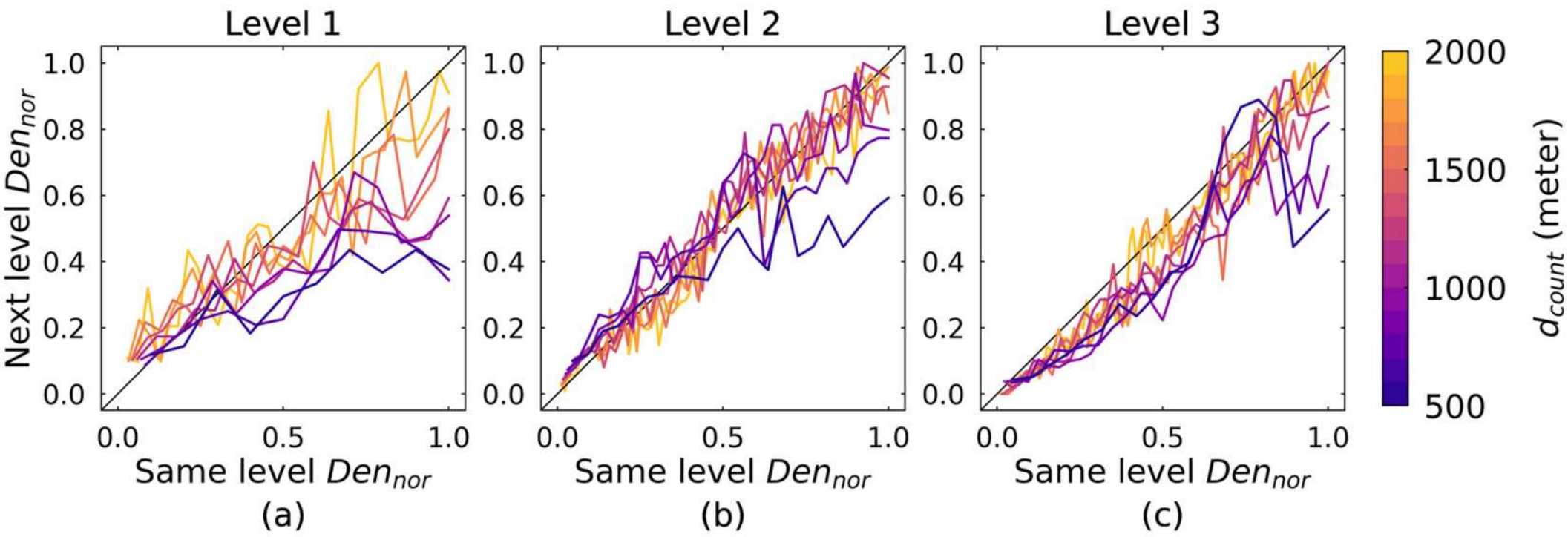

**Figure D2.** Mean curves of $(Den_{nor}(x, A, d_{count}|A), Den_{nor}(x, B, d_{count}|A))$ with different $d_{count}$ for POI local hotspots in Wuhan.

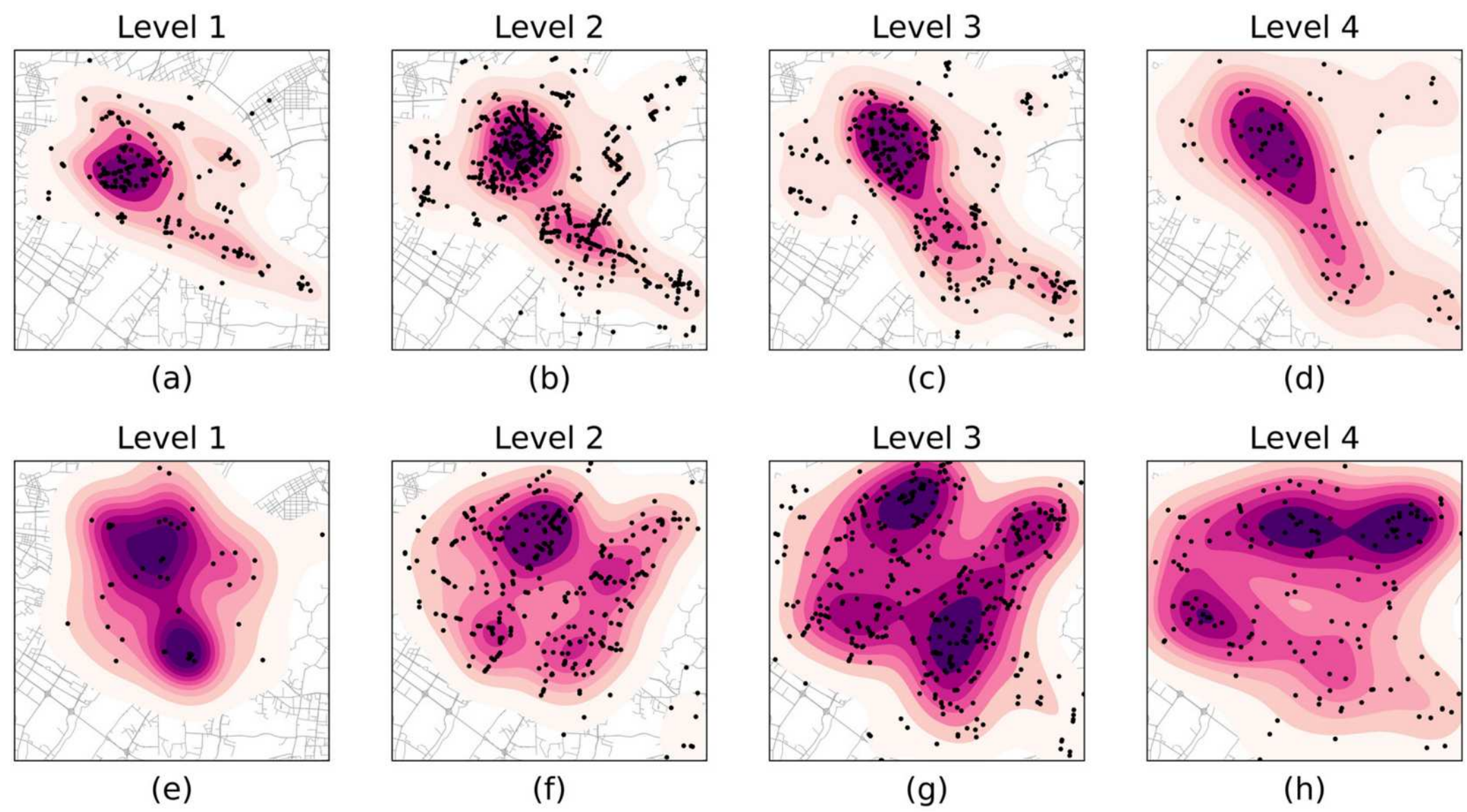

**Figure D3.** (a)–(d) are the spatial distribution of taxi pick-up hotspots in Wuhan within 200 m of the POI hotspot, and (e–h) are the cases beyond 200 m.